\documentclass{nature2}
\usepackage{graphicx}

\usepackage[english]{babel}
\usepackage[utf8x]{inputenc}
\usepackage[T1]{fontenc}
\usepackage{caption}

\usepackage[a4paper,top=3cm,bottom=2cm,left=2cm,right=2cm,marginparwidth=1.75cm]{geometry}

\usepackage{amsmath}
\usepackage[colorinlistoftodos]{todonotes}
\def \bea{\begin{eqnarray}}
\def \eea{\end{eqnarray}}
\def \be{\begin{equation}}
\def \ee{\end{equation}}

\def\({\left(} \def\){\right)}
\title{Pair formation in insect swarms driven by adaptive long-range interactions}
\author{Dan Gorbonos$^{1,*}$, James G. Puckett$^2$, Kasper van der Vaart$^3$, Michael Sinhuber$^{3,\dagger}$, Nicholas T. Ouellette$^{3}$ and Nir S. Gov$^{1,\ddagger}$}

\begin{document}
\maketitle
\begin{affiliations}
 \item Department of Chemical and Biological Physics, Weizmann Institute, Rehovot, Israel
 \item Department of Physics, Gettysburg College, Gettsyburg, Pennsylvania 17325, USA
 \item Department of Civil and Environmental Engineering, Stanford University, Stanford, California 94305, USA
 \end{affiliations}

\begin{abstract}
In swarms of flying insects, the motions of individuals are largely uncoordinated with those of their neighbors, unlike the highly ordered motion of bird flocks. However, it has been observed that insects may transiently form pairs with synchronized relative motion while moving through the swarm. The origin of this phenomenon remains an open question. In particular, it is not known if pairing is a new behavioral process or whether it is a natural byproduct of typical swarming behavior. Here, using an ``adaptive-gravity'' model that proposes that insects interact via long-range gravity-like acoustic attractions that are modulated by the total background sound (via ``adaptivity'' or fold-change detection) and that reproduces measured features of real swarms, we show that pair formation can indeed occur without the introduction of additional behavioral rules. In the model, pairs form robustly whenever two insects happen to move together from the center of the swarm (where the background sound is high) toward the swarm periphery (where the background sound is low). Due to adaptivity, the attraction between the pair increases as the background sound decreases, thereby forming a bound state since their relative kinetic energy is smaller than their pair-potential energy. When the pair moves into regions of high background sound, however, the process is reversed and the pair may break up. Our results suggest that pairing should appear generally in biological systems with long-range attraction and adaptive sensing, such as during chemotaxis-driven cellular swarming.
\end{abstract}

{\footnotesize\thanks{$^*$ Present address: Max Planck Institute of Animal Behavior, Dept. of Collective Behavior, Universitätsstraße 10, 78464 Konstanz, Germany.}}

{\footnotesize\thanks{$^\dagger$ Present address: Department of Physics, Carl von Ossietzky Universität, 26111 Oldenburg, Germany.}}

{\footnotesize\thanks{$^\ddagger$ E-mail: nir.gov@weizmann.ac.il}}

\section{Introduction}

Swarms are a form of collective animal behavior that have caught the attention of physicists as self-organized non-equilibrium systems that remain cohesive yet exhibit no clear order parameter \cite{vicsek2012collective}, as opposed to ``flocking'' behavior \cite{ballerini2008b}. Such behavior is observed in a variety of species, including fish \cite{lopez2012behavioural,tunstrom2013collective}, bats \cite{cvikel2015bats}, and flying insects \cite{giardina2014plos}. Theoretical models proposed to describe this collective phase often assume short-range (or near-neighbor) interactions \cite{couzin2002,ballerini2008} that contain a fine balance between attraction, repulsion, a tendency of the individuals to align their motion with that of their neighbors, and the effects of noise \cite{canizo2010collective}. It has even been suggested that certain insect swarms may be finely tuned to be poised close to a critical point where global alignment of motion would commence \cite{giardina2014prl}.

In an alternative framework \cite{gorbonos2016long}, we recently proposed that the interactions between flying insects (midges, in this case) are mediated by acoustics due to the sound they emit while flying, which gives rise to long-range power-law interactions. Furthermore, we suggested that the interactions are attractive, so that individuals tend on average to accelerate towards each other in proportion to the intensity of the sound received. For pure acoustics, the functional form of this acceleration is similar to gravity (that is, proportional to $r^{-2}$), although similar behavior arises even if the exponent has a different value \cite{gorbonos2017stable}. An additional, and crucial, component of this model is adaptivity, common to most sensory systems in biology \cite{shoval2010}, whereby the sensitivity of the
midges to the received sound drops when there is a strong background sound. Exact adaptation means that the steady-state output is
independent of the steady-state level of input, which is part of a
fold-change detection mechanism \cite{shoval2010}. It was shown in Ref.~\cite{gorbonos2016long} that this ``adaptive-gravity'' model reproduces many steady-state features of midge swarms, such as the observed reduction in the average accelerations of the midges towards the swarm center in larger swarms \cite{kelley2013}. More recently the model was shown to account for the observed mass and velocity profiles within the swarms \cite{gorbonos2020similarities}.

\begin{figure}
\centering
\includegraphics[width=1\linewidth]{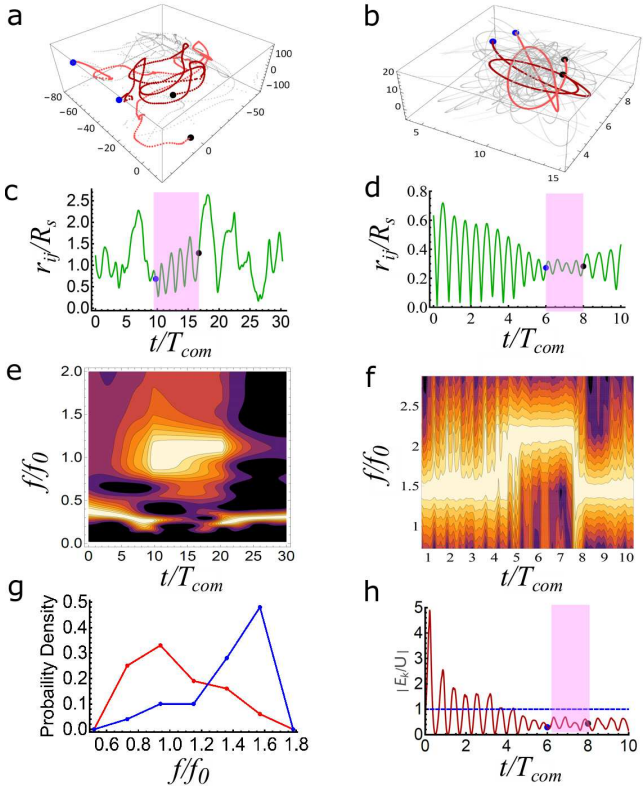}
\caption{\label{Figure1}  Pairs in laboratory observations of midge swarms (a,c,e) and simulations of the ALRI model (b,d,f).
(a) Trajectories of two midges in a laboratory swarm that exhibited pairing (defined as in ref.~\cite{puckett2015}). The midges were identified as belonging to a pair between the blue and black points. Paired parts of the trajectories are colored in red, while unpaired parts are in grey. Distances are in mm.
(b) A pair as identified in a simulation of the ALRI. Symbols and colors are the same as in (a). Distances are in simulation unit length.
(c,d) Distance between the members of the laboratory pair (c) and simulated pair (d) as a function of time. The blue and the black points correspond to the same points in (a,b), and the shaded region shows the period when the two individuals are paired. The distance is normalized by the swarm size $R_s$, which is defined as the mean distance of a midge from the center of mass of the swarm. Time is normalized by the typical orbit time around the center of mass, defined as $T_{com}=4\,R_s/\bar{v}$ where $\bar{v}$ is the mean midge speed (see Supplementary Material).
(e) Time-frequency analysis of the motion of the laboratory pair using a continuous wavelet transform, as explained in ref.~\cite{puckett2015}. Frequencies are normalized by the typical frequency $f_0=1/T_{com}$ and the amplitude by the time-resolved peak power $F_{max}(t)$. Higher powers are represented with brighter colors. During pairing, the peak power shifts to higher frequencies. (f) Time-frequency analysis of the motion of the simulated pair using a windowed Fourier transform with a window length of $0.14T_{com}$ (see Supplementary Material, section 2). The behavior is similar to what is seen in the laboratory pair.
(g) Probability density function (PDF) of the peak frequency of motion in the simulation during pairing (blue) and independent motion (red), where pairing is defined by the ratio of kinetic to potential energy being less than one. The increase in frequency during pairing seen in the example in (f) is statistically robust. See supplementary material for more details. 
(h) The ratio of kinetic to potential energy for the example shown in (b). Note the decrease in the energy ratio during pairing, which is defined by having $|E_K/U|<1$ when averaged over $T_{com}$.}
\end{figure}

In addition to steady-state features, recent observations have found evidence for the dynamic formation of synchronized pairs of midges, which typically oscillate with respect to each other at a higher-than-normal frequency and maintain a small distance between them while they move together through the swarm (Fig.~\ref{Figure1}a) \cite{puckett2015}. Pairs were identified in the laboratory swarms via an increase in the frequency of mutual oscillation of two midges (Fig.~\ref{Figure1}c,e) that persisted longer than a threshold time \cite{puckett2015}. During pairing, the amplitude of the relative oscillations of the pair also diminished (Fig.~\ref{Figure1}c), but remained much larger than the distance where  midges might accelerate away from each other to avoid collision. However, no mechanism for this phenomenon was proposed. In particular, it is not known if pairing is a result of additional behavioral rules or whether it can arise naturally as a passive byproduct of swarming. Here we report that the same model of adaptive long-range interactions (ALRI) \cite{gorbonos2016long} that captures many steady-state features of swarms indeed produces pairing without any modifications. Thus, pairing can be viewed as an emergent phenomenon and a natural outcome of the same interactions that lead to swarm formation.


Pairing is a rare event in systems with attractive long-range interactions in the absence of adaptivity (such as classical gravity). Such systems are described by a Hamiltonian, and thus conserve energy and momentum. Due to momentum conservation, the capture of two particles to form an orbiting pair must involve a third particle that will remove the excess momentum. Such situations are highly unlikely to occur, and indeed under classical gravity stellar pairs rarely form~\cite{bodenheimer2011principles}. Adaptivity, however, means that the system does not obey energy or momentum conservation \cite{gorbonos2016long}, and the dynamics is not limited by these constraints. This has significant consequences for pairing, as we show below.

\section{ALRI model}
The basic equation of motion of the ALRI model gives the effective force on midge $i$ due to the sum over all the other midges $j$ as
\begin{equation}
\vec{F}^{i}_{\mbox{\scriptsize eff}}=C\sum_{j}\hat{r}_{ij}\frac{1}{|\vec{r}_i-\vec{r}_j|^2+\epsilon^2}\left(\frac{R_{\mbox{\scriptsize ad}}^{-2}}{R_{\mbox{\scriptsize ad}}^{-2}+\sum_{k}(|\vec{r}_i-\vec{r}_k|^2+\epsilon^2)^{-1}}\right) \label{feffmany},
\end{equation}
where $\vec{r}_i$ is the position vector for midge $i$, $\hat{r}_{ij}$ is the unit vector pointing from midge $i$ to midge $j$, $C$ is a constant with dimensions of $mass \cdot length^3/time^{2}$, $\epsilon$ is a constant with units of length, and $R_{\mbox{\scriptsize ad}}$ is the length scale over which adaptivity occurs. For $r_{ij}\gg \sqrt{N}R_{\mbox{\scriptsize ad}}$ where $N$ is the number of midges in the swarm (that is, when the distance between a pair of midges far exceeds the range of adaptivity), the effective force reduces to a purely gravitational interaction. 
For comparison, we also considered an $\epsilon$-gravity model, which is non-adaptive, classical gravity that is softened to prevent runaway accelerations that produce slingshots that break up the swarm too quickly \cite{gorbonos2020similarities}. In $\epsilon$-gravity, the effective force on a midge is given by
\begin{equation}
\vec{F}^{i}_{\mbox{\scriptsize eff}}=C\sum_{j}\hat{r}_{ij}\frac{1}{|\vec{r}_i-\vec{r}_j|^2+\epsilon^2} \label{epsilon}.
\end{equation}
Note that we consider here point particles (so that they do not collide), without any short-range repulsion; this assumption does not affect the overall trajectories\cite{gorbonos2016long}. We use $\epsilon^2=15$ and $C=1$ throughout.

We looked for evidence of pairing in both the ALRI and $\epsilon$-gravity models by simulating their behavior. For details on the
simulation technique and initial conditions \cite{gorbonos2020similarities}, see Section 1 of the Supplementary Material.

In classical gravity, as in any Hamiltonian system, a bound pair of objects is defined by having a kinetic energy with respect to each other that is smaller than the potential energy between them. For the ALRI model in a swarm of $N>2$ particles, we do not have a well-defined potential energy \cite{gorbonos2016long}. However, we may use as an approximation the expression for a pair of particles, assuming that the rest of the swarm contributes to leading order only a uniform background sound. Integrating the force acting on the midges (Eq.~\ref{feffmany}), we can calculate the effective two-body potential to be 
\begin{equation}
U_{\mbox{\scriptsize pair}}(r)=\frac{C}{\gamma\,\sqrt{R_{\mbox{\scriptsize ad}}^2+\epsilon^2\gamma^2}}\left(\arctan{\left(\frac{\gamma\,r}{\sqrt{R_{\mbox{\scriptsize ad}}^2+\epsilon^2\gamma^2}}\right)}-\frac{\pi}{2}\right),\nonumber
\end{equation}
where $r\equiv|\vec{r}_1-\vec{r}_2|$ and we approximate $\gamma$ as a constant for a pair (see Supplementary Material). We thus take $\gamma$ to be the average of $\gamma(\vec{r}_1)$ and $\gamma(\vec{r}_2)$, so that
\begin{equation}
\gamma\equiv\frac{1}{2}\left[\gamma(\vec{r}_1)+\gamma(\vec{r}_2)\right]
\end{equation}
where
\begin{equation}
\label{gamma}
\gamma(\vec{r}_i) \equiv \sqrt{1+R_{ad}^2\,I_{background}(\vec{r}_i)} \qquad i=1,2
\end{equation}
and
\begin{equation}
I_{\mbox{\scriptsize background}}(\vec{r}_i) = \sum_{j=3}^{n}\frac{1}{|\vec{r}_i-\vec{r}_{j}|^{2}+\epsilon^2}
\end{equation}
is the parameter that quantifies the background sound at the location of the pair and is independent of the distance $r$ between the pair members. Adaptivity weakens the interactions and therefore slows down the simulated particles (Fig.~S5), and so we normalize all times by the typical orbit time across the swarm $T_{com}=4\,R_s/\bar{v}$, where $\bar{v}$ is the mean midge speed (see Supplementary Material) and $R_s$ is the swarm size, defined as the mean distance of a midge from the center of mass of the swarm.

\section{Results}
\subsection{Pairs in the ALRI model.}

We use this approximate potential energy to define bound pairs in simulations of the ALRI model as pairs whose ratio of relative kinetic and potential energies is less than one: $|E_K/U|<1$ (Fig.~\ref{Figure1}h), averaged over a duration of $T_{com}$. While we cannot use the energy ratio criterion to analyze data from real midge swarms (since we do not know the quantitative strength of the interactions), we can compare other features of pairs between observational data and simulations. The common features of increased frequency (Fig.~\ref{Figure1}e,f) and diminished amplitude (Fig.~\ref{Figure1}c,d) are found for most pairs in both observations (with pairs defined as in ref.~\cite{puckett2015}) and simulations (with pairs defined using the energy ratio, Fig.~\ref{Figure1}h). In Fig.~\ref{Figure1}g we show that bound pairs in simulations are highly likely to exhibit the higher-than-normal frequency mutual oscillations that were used to identify pairs in the observational data (for details of the frequency calculation see Supplementary Materials, Section 2). These similarities suggest that the mechanism driving pairing in the ALRI model may also be present in real swarms.

\begin{figure}
\centering
\includegraphics[width=1\linewidth]{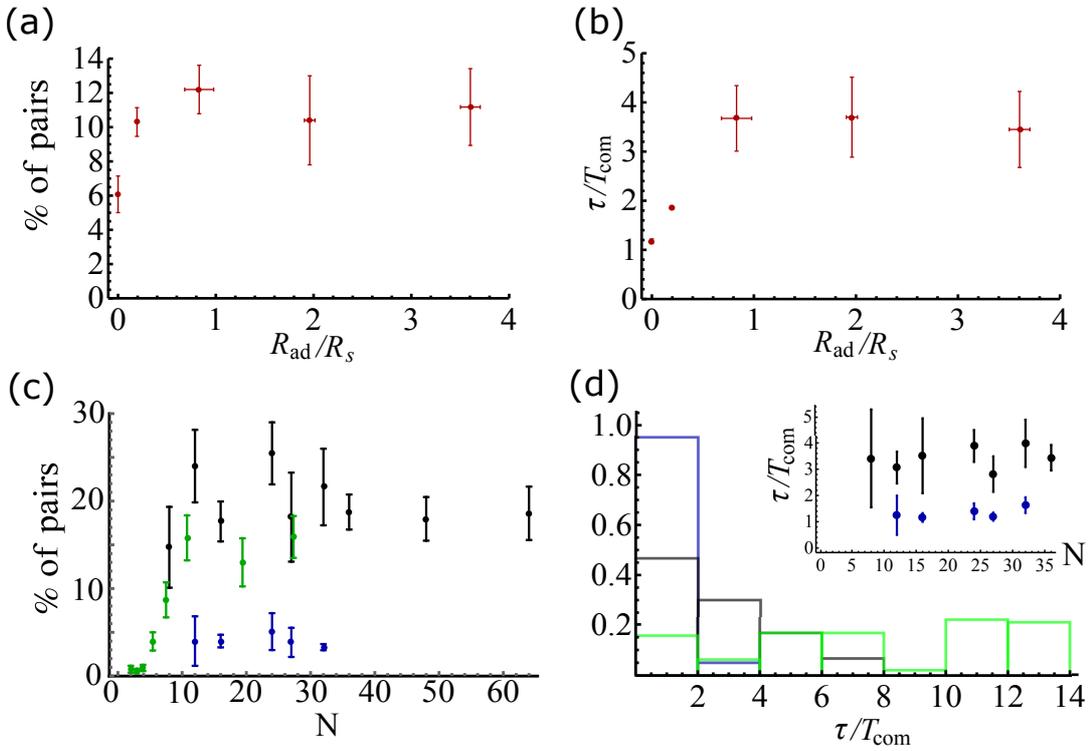}
\caption{\label{Figure2}
Pairing statistics.
(a) The mean percentage of time a midge spends in a pair ("fraction of pairs"), in a simulation of the ALRI model with $N=30$ midges as a function of $R_{ad}/R_{s}$.
(b) The mean lifetime of a pair ($\tau$) in normalized time units as a function of $R_{ad}/R_{s}$ for the same data as in (a).
(c) The fraction of pairs as a function of the number of members of the swarm $N$ for both simulations with $R_{ad}=10$ (black) and laboratory swarm observations (green). In blue we give the results from simulations of the non-adaptive $\epsilon$-gravity system. 
(d) PDF of the lifetime of pairs without adaptivity in blue ($R_{ad}=0$,$R_{s}=4.9$,$N=36$), with adaptivity in black ($R_{ad}=50$,$R_{s}=5.1$,$N=36$), and for laboratory swarm measurements in green ($N=21$). All are in normalized time units. In the inset we show that the mean lifetime of pairs is independent of the swarm size, from simulations with adaptivity in black and without adaptivity in blue.}
\end{figure}

Now that we have shown that pairing exists in the ALRI, we can quantify some of its features. In Fig.~\ref{Figure2}a we plot the fraction of time a midge spends in a pair (that is, the ``fraction of pairs'') as a function of the length scale of adaptivity $R_{\mbox{\scriptsize ad}}$ for a simulated swarm of size $N=30$. The limit of $R_{\mbox{\scriptsize ad}}\rightarrow0$ corresponds to the non-adaptive $\epsilon$-gravity system. Although there are some pairs found in this limit, they are highly transient and their mean lifetime is small (Fig.~\ref{Figure2}b). In Fig.~\ref{Figure2}c we plot the fraction of pairs as function of the swarm size for a fixed $R_{\mbox{\scriptsize ad}}=10$. In our previous analysis \cite{gorbonos2016long}, we found that real swarms lie in the strong adaptivity regime (where $R_{\mbox{\scriptsize ad}}\gg R_{\mbox{\scriptsize s}}$). It is therefore highly satisfying that in this limit, where there is no free parameter in the model, we find that the fraction of pairs is similar to the observations \cite{puckett2015} (Fig.~\ref{Figure2}c). 
The distribution of pair lifetimes in the simulations (Fig.~\ref{Figure2}d), shows that long-lived pairs do not arise in $\epsilon$-gravity.  However, ALRI is consistent with measurements from laboratory swarms, as we find there are pairs that survive for many orbits.
Note that due to potential reconstruction difficulties in the observations leading to broken trajectories, the lifetimes of the pairs from the laboratory observations should be considered to be a lower bound.

Thus, the ALRI model naturally exhibits pairing, and these pairs bear a number of similarities to those observed in real swarms. It is thus natural to ask what features of the ALRI model produce this pairing, and whether these key features are likely to be present in real swarms. The critical component appears to be adaptivity. The difference in dynamics between the ALRI and $\epsilon$-gravity models is strikingly apparent simply from watching movies of the two simulations (see Supplementary Movie 1 and Movie 2): in the ALRI case, pairs of particles are easily detectable by eye (Fig.~\ref{Figure1}b,d), while for $\epsilon$-gravity no pairs are evident. Other features, such as noise or imperfect behavioral response, appear to be less important. In our simulations, the midges are not described as noisy self-propelled particles as is common in classical models of collective behavior \cite{vicsek1995}; rather, they simply move inertially according to the effective forces (Eq.~\ref{feffmany}) that they feel from the other midges. Effective stochasticity in the trajectories arises from the complexity of Hamiltonian many-body dynamics (see for example \cite{meyer2008introduction}). In addition to this stochasticity, the trajectories of real midges seem to be affected by additional sources of high-frequency and small-amplitude noise (compare, for example, Figs.~\ref{Figure1}a,b, and see Figs.~S3,S4, Supplementary Material Section 3), which does not seem to qualitatively change the large scale dynamics of the midges. Since the ALRI is a minimal model, there are certainly additional effects in real swarms that it does not capture \cite{gorbonos2020similarities}. Nevertheless, it does not appear that these other effects, although important for determining the details in real swarms, are required to obtain pairing.

\subsection{Pairing mechanism in the ALRI model.}

Due to adaptivity (Eq.~\ref{feffmany}), it is clear that when a midge is close to another within a pair, the strong sound received from its partner acts to screen out the forces due to more distant midges in the swarm \cite{gorbonos2016long}. However, this observation does not explain how adaptivity induces the capture of two midges into a bound pair.

\begin{figure}
\centering
\includegraphics[width=0.8\linewidth]{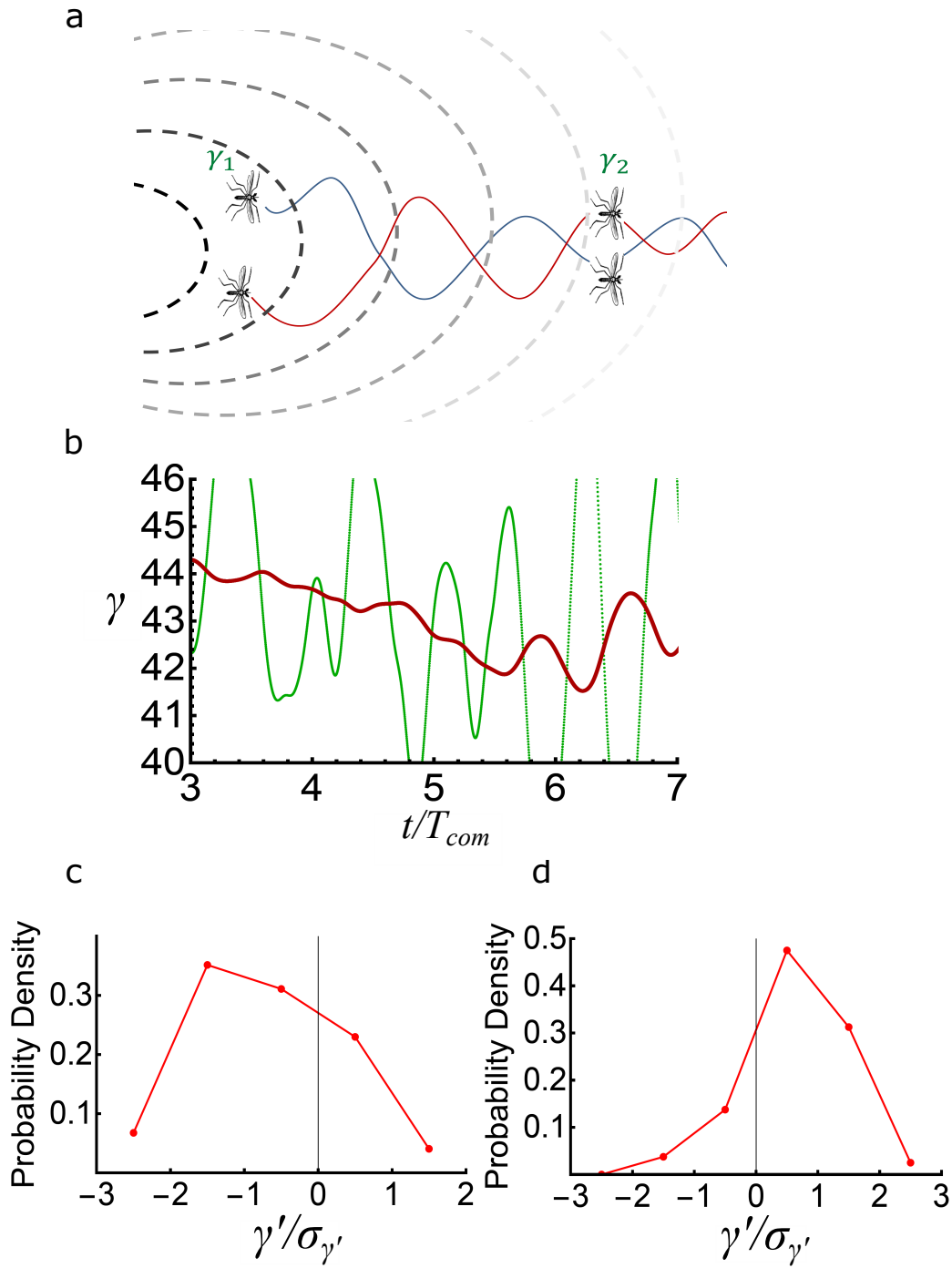}
\caption{\label{Figure3}
Pair formation.
(a) Illustration of the proposed pair formation mechanism. When two interacting midges leave the dense region of the swarm (darker dashed lines), where the background sound $\gamma$ is high, and move to a lower density region (such that $\gamma_1>\gamma_2$), the mutual pull between them becomes stronger, their orbit gets tighter, and they become bound.
(b) The background sound $\gamma$ along the path of the simulated pair shown in Fig.~\ref{Figure1}b, showing the raw sound (green) and the signal averaged over $T_{com}$ (magenta). When the background sound $\gamma$ is decreasing (for $t/T_{\mathrm{com}} \in [3,6]$), the amplitude of the pair oscillations also decreases (as in Fig.~\ref{Figure1}d), and finally the pair is formed (Fig.~\ref{Figure1}h, $t/T_{\mathrm{com}}>5$).
(c) PDF of the gradient of the dimensionless background sound at the time when the pair is formed. Data are taken from 250 cases from simulated swarms with $R_{ad}=50$, $R_{s}=5.1$, and $N=36$ where the mean energy ratio $E_k/U$ was lower than 1 for a time segment of at least $T_{com}$. The background sound gradient ($\gamma'$) tends to be negative during pair formation, in agreement with our theoretical predictions. The mean value of the dimensionless background sound gradient is $-0.65\sigma$. (d) The same statistics as in (c) for pair dissociations. Here the mean gradient is positive, with a mean value of $0.56\sigma$.}
\end{figure}

In Fig.~(\ref{Figure3}a) we show schematically how this process happens. Suppose two midges are close to each other in the inner part of the swarm, where the background sound level is high and therefore their attraction toward each other is weak. If they happen to be moving together away from the swarm center, they will experience decreasing background sound levels, resulting in stronger mutual attraction. These two midges initially moved toward each other in a regime of weak mutual attraction (high background sound),  gaining little kinetic energy in the process, but now find themselves in a regime of strong attraction (low background sound) that binds them together as a pair. A mathematical analysis of this process, whereby a decrease in the background sound leads to a tightening of the orbits of the pair, is given in Section 6 of the Supplementary Material. And indeed, the radial distribution of pair-formation events is found in simulations to be concentrated in the high density region of the swarm (Fig.~S11), as there the particles are closest to each other and are likely to be moving from high to low background sound. The mechanism of pair formation in the ALRI model is therefore a many-body effect (since the two midges in this model are still described by Hamiltonian dynamics), but unlike capture in non-adaptive gravity, which hardly ever occurs, the production of pairs happens robustly. This difference is further illustrated in Section 7 of the Supplementary Material (Fig.~S10).

In the ALRI model, the process of pair formation when moving from high to low background sound is reversed when pairs move from low to high background sound. Since the system obeys time-reversal symmetry, this reverse process acts to break up the pairs (illustrated in Fig.~S10). Note that in the regime of strong adaptivity ($R_{\mbox{\scriptsize ad}}\gg R_s$), the pairing behavior should not be strongly dependent on the exponent of the power-law of the long-range interaction \cite{gorbonos2017stable}. In the simulations we also find triplets that form, though at significantly lower proportions (Figs.~S6,S7, Supplementary Section 5).

We can test the validity of this proposed mechanism by computing the gradient of the background sound along the trajectory of the pair at the time of pair formation. In other words, we calculate $\gamma'=d\gamma/dr_{pair}$, where the background sound $\gamma$ is defined in Eqs.~\ref{gamma} and the gradient is calculated along the path of the pair's center of mass $r_{pair}$. A specific example is shown in Fig.~\ref{Figure3}b for the simulated pair shown in Fig.~\ref{Figure1}b,d. We calculated the statistics of $\gamma'$ at the time of pair formation (Fig.~\ref{Figure3}c). There is a clear asymmetry in the distribution of the background sound gradient, with a skewness towards decreasing values along the pair's trajectory at the time of pair formation. Similarly, when pairs break, the mechanism in the model is the increasing background sound (Fig.~\ref{Figure3}d).

\section{Discussion}
We have thus demonstrated that additional behavioral rules are not necessary to drive the formation of pairs in midge swarms, and that pairing (like swarming itself) can be viewed as an emergent phenomenon. Note, however, that our theory does not tell us about the biological function of pairing, since pairing and swarming are inexorably linked in this model. We find that the key ingredients that give rise to pairing are long-range interactions and adaptivity. These features appear in many different contexts and on different scales, including in cellular swarming \cite{daniels2004quorum} and insect swarming driven by chemical communication \cite{slessor2005pheromone}. Our work thus argues that pairing should be a general feature that emerges in biological collective systems that have long-range attractive interactions with adaptivity.

\noindent \textbf{Acknowledgments.} The research at Stanford was sponsored by the Army Research Laboratory and accomplished under grant no.~W911NF-16-1-0185. The views and conclusions in this document are those of the authors and should not be interpreted as representing the official policies, either expressed or implied, of the Army Research Laboratory or the U.S. government. K.v.V. acknowledges support from an Early Postdoc Mobility fellowship from the Swiss National Science Foundation, and M.S. acknowledges support from the Deutsche Forschungsgemeinschaft under grant no.~396632606. N.S.G is the incumbent of the Lee and William Abramowitz Professorial Chair of Biophysics. This work is made possible through the historic generosity of the Perlman family.


\bibliographystyle{naturemag}
\bibliography{acoustics_1}

\end{document}


\title{Supplementary Information: ``Pair formation in insect swarms driven by adaptive long-range interactions''}

\author{Dan Gorbonos$^{1,*}$, James G. Puckett$^2$, Kasper van der Vaart$^3$, Michael Sinhuber$^{3,\dagger}$, Nicholas T. Ouellette$^{3}$ and Nir S. Gov$^{1,\ddagger}$}


\affiliation{$^1$ Department of Chemical Physics, The Weizmann Institute of Science, P.O. Box 26, Rehovot, Israel 76100 \\
$^2$ Department of Physics, Gettysburg College, Gettsyburg, Pennsylvania 17325, USA \\
$^3$ Department of Civil and Environmental Engineering, Stanford University, Stanford, California 94305, USA\\
\\
\\
$^*$ Present address: Max Planck Institute of Animal Behavior, Dept. of Collective Behavior, Universitätsstraße 10, 78464 Konstanz, Germany.\\
$^\dagger$ Present address: Department of Physics, Carl von Ossietzky Universität, 26111 Oldenburg, Germany.\\
$^\ddagger$ E-mail: nir.gov@weizmann.ac.il
}

\maketitle

\setcounter{equation}{0}  
\setcounter{figure}{0}

\section{The simulation}
\label{simulation}

We used a program that was developed to compute the time evolution of $N$ particles under the influence of their mutual self-gravity. To focus on the effects of the proposed adaptive-gravity interactions, we did not include in the simulation any explicit noise terms. The original program was written by S. J. Aarseth for ``regular'' gravity (without adaptivity) and appeared in~\cite{binney2011galactic} appendix 4.B. A complete description of the numerical method is given in~\cite{brackbill1985multiple} p. 377. The acceleration of each mass is computed by the direct summation of the forces due to the other $N-1$ bodies
according to Eq.~(1) . The program was designed to work efficiently with up to 50 masses, which is similar to the number of midges in swarms (most of the swarms consist of up to 50 individuals). The particles in the simulation show a strong tendency to evaporate because they develop high accelerations in close encounters~\cite{PhysRevResearch.2.013271}. In order to avoid close encounters, it is common to introduce a softening parameter $\epsilon$ to the gravitational force that appears in Eq.~(1).

The initial conditions in the simulations are as follows.
 We chose initial positions that give us approximately the same kinetic energy per individual in the case of simulations with adaptivity (within about 10\%).  Without adaptivity, we could not change significantly the kinetic energy by varying the initial conditions, and so it was determined mainly by the number of particles in the simulation.
The initial velocities are all zero. So the initial kinetic energy is zero.  Numerical experiments with different values of $\epsilon$ showed that in order to avoid a quick evaporation of the swarm, one has to take $\epsilon$ to be of the order of the initial distances between the particles~\cite{PhysRevResearch.2.013271}. Our simulations were performed with $\epsilon^2=15$ and $C=1$.

\begin{figure}[htb]
     \includegraphics[width=0.7\linewidth]{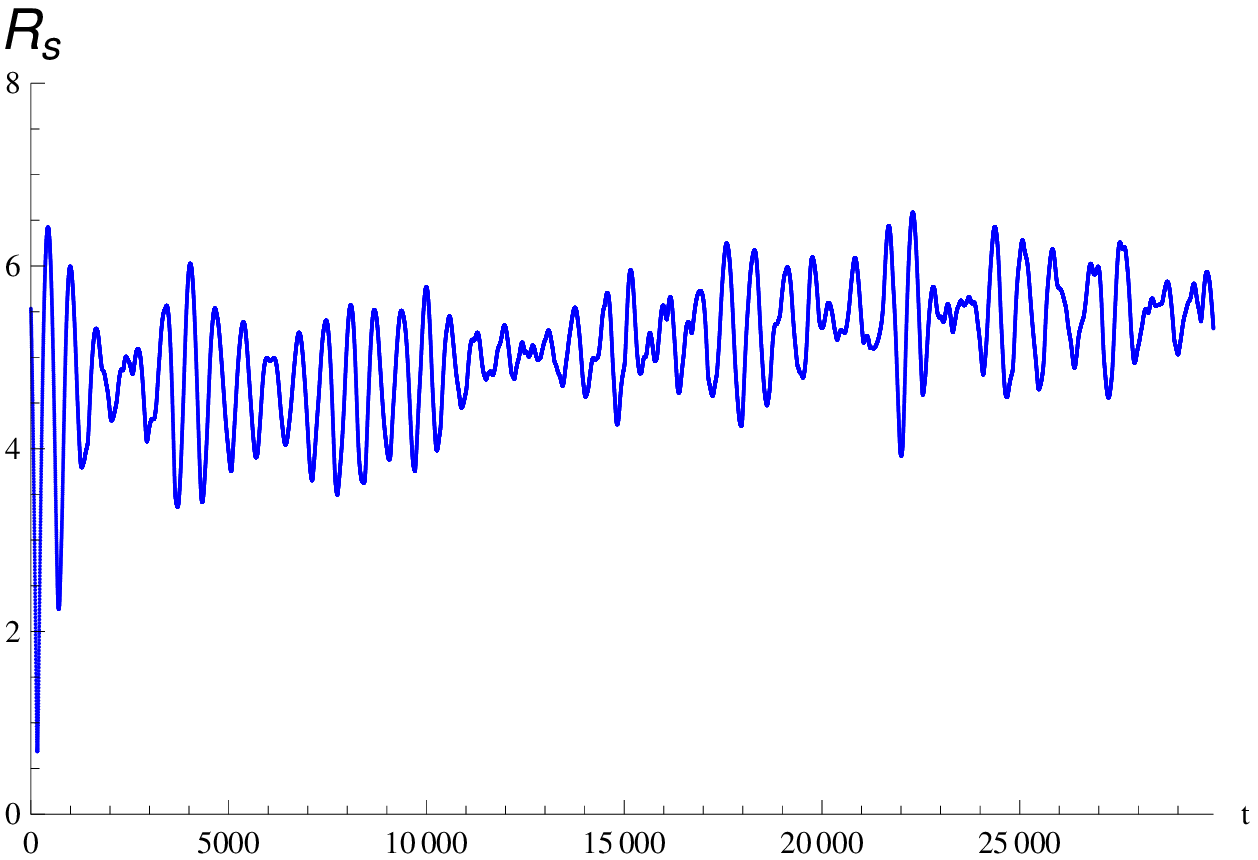}
     \vspace{1.5cm}
\caption{ The size of the swarm $R_s$ as a function of the simulation time for a swarm of 36 midges with $R_{ad}=50$. The oscillation period is approximately
$T_{com}=4\,R_s/\bar{v}$, where $R_{s}$ is the swarm size and $\bar{v}$ is the mean speed of a midge.
}
\label{RsExample}
\end{figure}

Fig.~\ref{RsExample} shows the size of the swarm as a function of time for a typical swarm. After the initial collapse that gives the system kinetic energy (remember that adaptivity is responsible for deviation from conservation of energy), the system goes through some large fluctuations and eventually enters a ``pseudo-stable'' mode in which the fluctuations are small relative to the size of the swarm $R_{s}$, which is defined as the mean distance of a midge from the center of the mass of the swarm. Still the swarm evaporates and $R_{s}$ slowly grows but in this regime it happens on a longer timescale than the typical time for an orbit around the center of mass, defined as $T_{com}=4\,R_s/\bar{v}$ where $\bar{v}$ is the mean speed for movement in the swarm. This is the relevant timescale for all the processes in the swarm including a period of oscillation of the size of the swarm itself as seen in Fig.~\ref{RsExample}.

In the example of Fig.~\ref{RsExample} $R_{s}\sim 4.6$ and $\bar{v}\sim 0.028$ for $t<10^4$, and then $T_{com}\sim 680 \ll 10^4$. For $10^4<t<2*10^4$ $R_{s} \sim 5.2$ and $\bar{v}\sim 0.03$, and then $T_{com}\sim 690 \ll 10^4$. So $R_{s}$ clearly increases as a function of time but very slowly relative to $T_{com}$.

%

\section{Frequency distribution, the windowed Fourier transform and the Wavelet transform}
\begin{figure}[htb]
\includegraphics[width=0.3\linewidth]{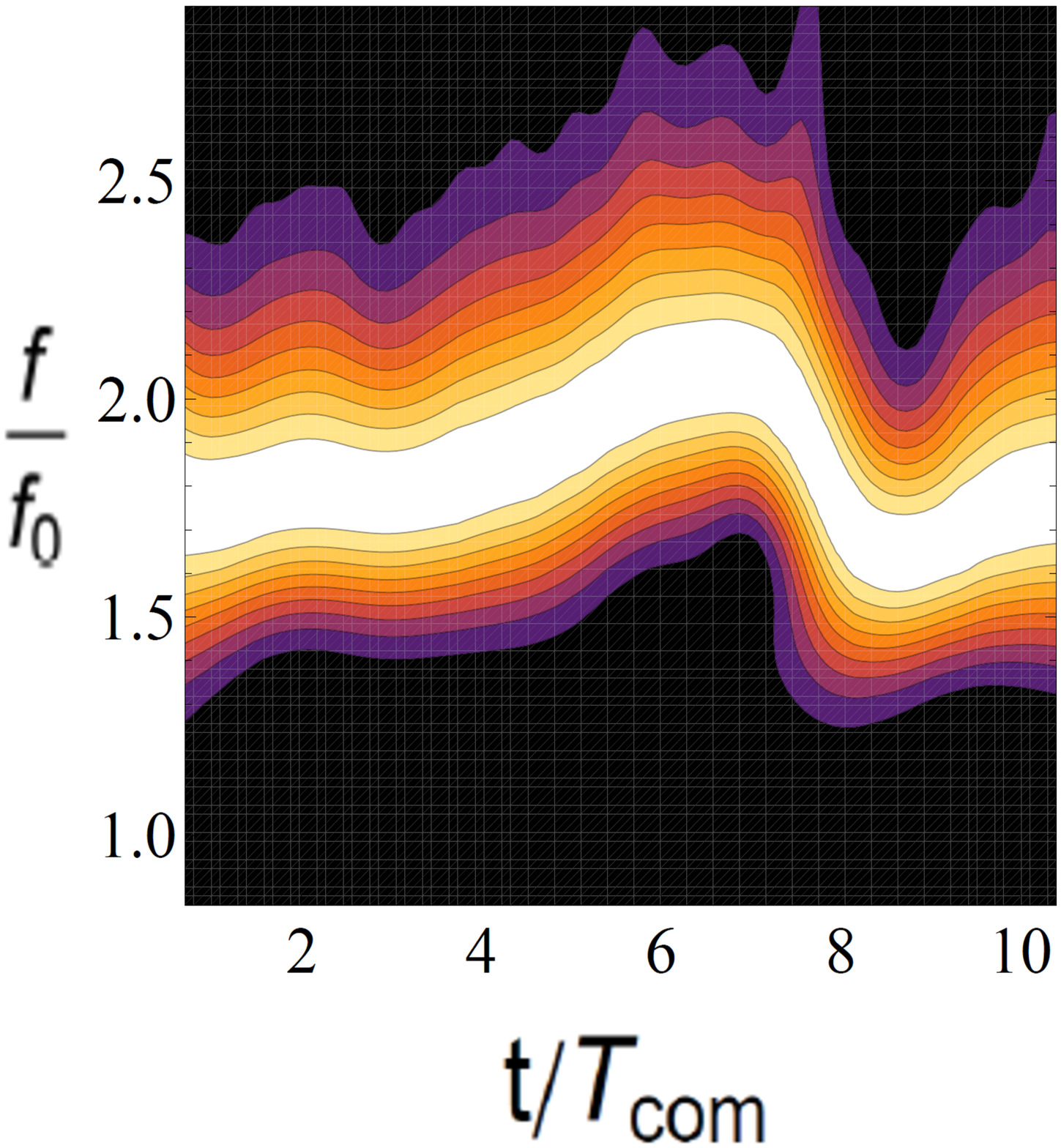}
\vspace{1.5cm}
\caption{The wavelet transform of the time series of the distance between the members of the pair of the simulation that appears in Fig.~1d. The result here is smoother than the windowed Fourier transform that appears in Fig.~1f and essentially contains the same information. }
\label{WaveletExample}
\end{figure}

For the spectrogram of Fig.~1f we used a windowed Fourier transform. At every instant of time, the Fourier coefficient was computed for a time segment whose width is $0.14T_{com}$, where $T_{com}=4\,R_{s}/\bar{v}$ where $\bar{v}$ is the mean speed for movement in the swarm and $R_s$ is the swarm radius (see previous section). This width was chosen since it appeared to be narrow enough in order to see the rapid changes in the frequency. The frequency was then identified according to the maximal Fourier component. We prefer to use the windowed Fourier transform for the simulation data and not the wavelet transform that was used for the data from laboratory observations (Fig.~1e) since the wavelet transform gives a more smoothed out and less distinct transition into the paired state. The trajectories in the simulations are much smoother than in the laboratory swarms, and the changes of the frequency of oscillation when the pair forms are much less sharp as compared to the changes of the frequency in laboratory swarms (see section 3). For comparison we give here in Fig.~\ref{WaveletExample} the wavelet transform of the same data computed according to the procedure that was described in~\cite{puckett2015} (and used to produce Fig.~1e).

In Fig.~1g we used a weaker localization in time than the windowed Fourier transform. For each time segment whose length is $T_{com}$ we took a segment whose length is $6\,T_{com}$ around it and computed the discrete Fourier transform for this segment. Again, the frequency was identified according to the maximal Fourier component. Such a length was necessary in order to catch most of the lower frequencies in the examples that were considered. We paid by losing some accuracy in the identification of the frequency with the ratio of kinetic to potential energies in the segment. Even with such a clear drawback there is a very clear correlation between the high frequencies (the criterion for pairs that was used for laboratory swarms in~\cite{puckett2015}) and the definition that was used here of a bound pair when averaged over a duration of $T_{com}$:

\be
\mid E_{k}/U_{\mbox{\scriptsize pair}} \mid <1.
\ee

\section{The difference between the trajectories in the laboratory and the simulation in terms of curvature}

\begin{figure}[htb]
     \includegraphics[width=0.9\linewidth]{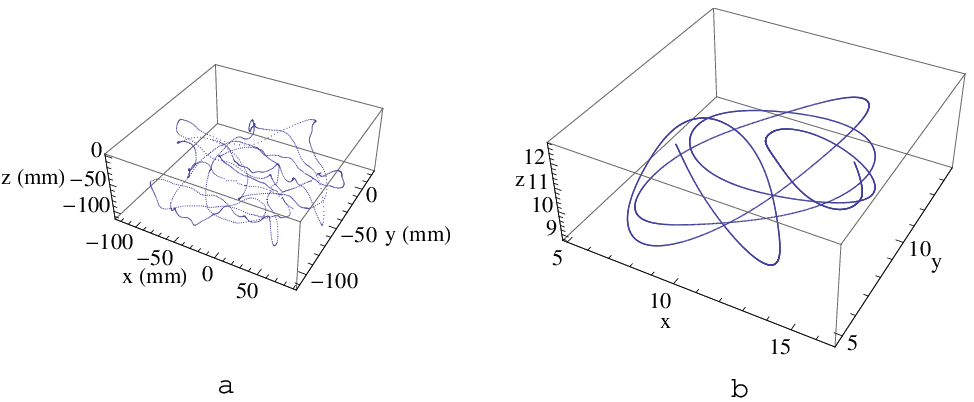}
     \vspace{1.5cm}
\caption{ (a) A typical trajectory reconstructed from the laboratory data. (b) A typical trajectory of a midge from the simulation. Note that the curve of the observed trajectory has more cusps and therefore higher fluctuations in curvature.
}
\label{TypicalCurves}
\end{figure}

\begin{figure}[htb]
     \includegraphics[width=0.7\linewidth]{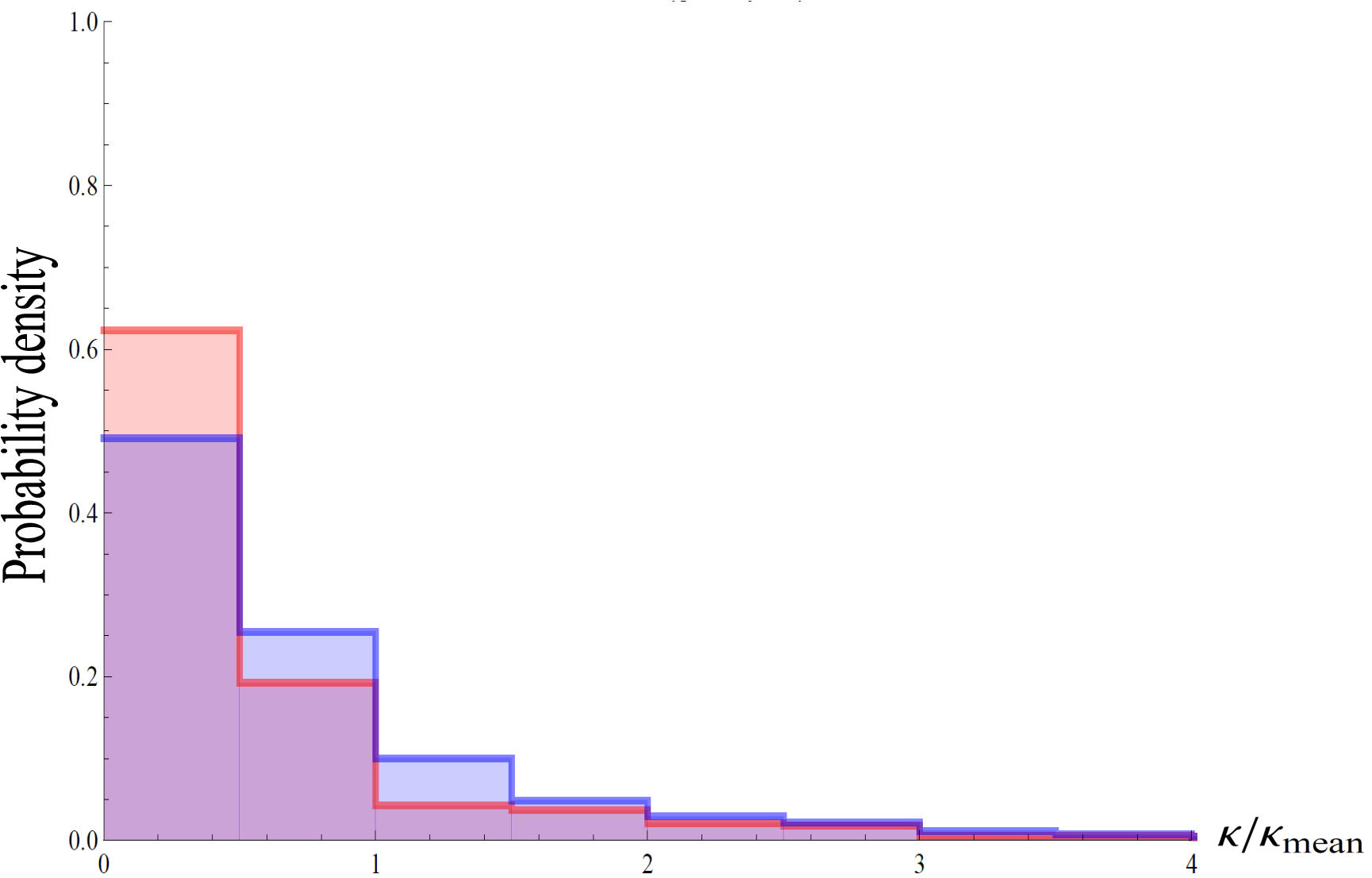}
     \vspace{1.5cm}
\caption{The distribution of curvature for the observed trajectories (blue) and the trajectories from the simulation (red). Both are normalized by the mean curvature.
}
\label{HistCurv}
\end{figure}

In Fig.~\ref{TypicalCurves} a typical trajectory from the laboratory is compared with a typical trajectory from the simulation.  The trajectory from the laboratory has more cusps and sharp edges. In order to quantify this, let us consider the distribution of the curvature on the trajectories. Taking three subsequent points on each trajectory $\vec{r}(t-\Delta t)$, $\vec{r}(t)$ and $\vec{r}(t+\Delta t)$ (where $\Delta t$ is a small time separation), we can construct the following sides of a triangle:
\bea
a&=&\left|\vec{r}(t+\Delta t)-\vec{r}(t)\right| \nonumber\\
b&=&\left|\vec{r}(t)-\vec{r}(t-\Delta t)\right| \nonumber\\
c&=&\left|\vec{r}(t+\Delta t)-\vec{r}(t-\Delta t)\right|.
\eea
Then, since the curvature $\kappa$ at a point is the inverse of the radius of the osculating circle at a point, using the sine rule and Heron's formula for the area of a triangle, one can write the following expression for the curvature:
\be
\kappa(t)=\frac{\sqrt{(a+b+c)(b+c-a)(a+c-b)(a+b-c)}}{a\,b\,c}.
\ee
In Fig.~\ref{HistCurv} we compare the distribution of the curvature along typical curves in the laboratory swarms and the simulation, normalized by the mean value of the curvature. We can see clearly that the trajectories of the midges in the laboratory swarms have more frequent higher values of curvature. This way we can quantify this property of cusps and sharp edges.

\section{The mean speed in the swarm as a function of the adaptivity length scale $R_{ad}$ (simulation)}
\begin{figure}[htb]
     \includegraphics[width=0.6\linewidth]{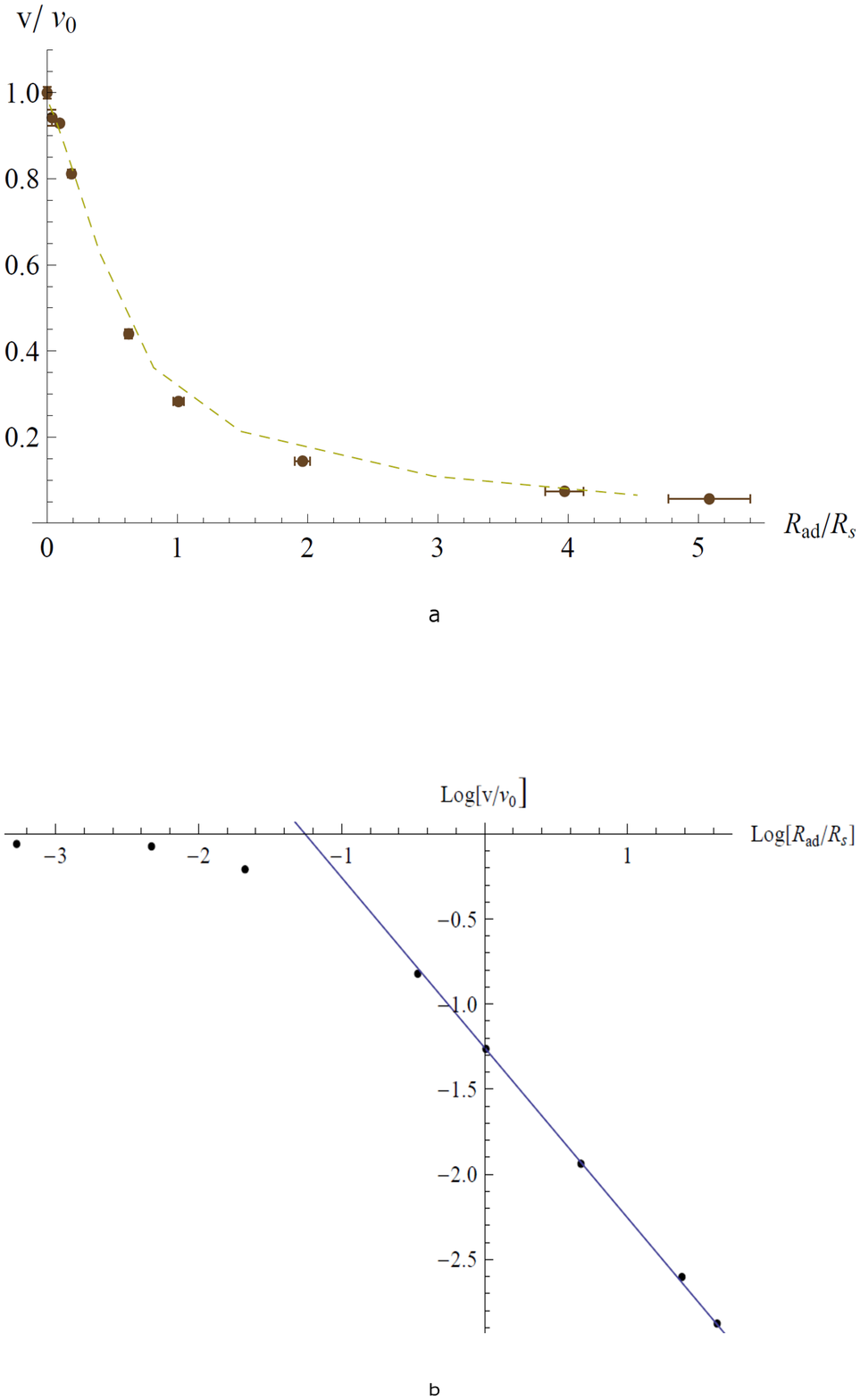}
     \vspace{1.5cm}
\caption{(a) The mean speed as a function of the strength of adaptivity $R_{ad}/R_{s}$ normalized by the maximal speed (when there is no adaptivity $R_{ad}=0$).
(b) The same as (a) plotted on logarithmic axes. For  $R_{ad}>R_{s}$, $\bar{v}\propto R_{ad}^{-1}$ which is the plotted line.
}
\label{Speed}
\end{figure}

As adaptivity becomes stronger, the mean speed of the midges becomes slower as shown in Fig.~\ref{Speed}a. In Fig.~\ref{Speed}b the same data is shown on logarithmic axes. In perfect adaptivity, when $R_{ad}>R_{s}$, the force $F \propto R_{ad}^{-2}$ and then the potential (and kinetic) energies
$U \sim E_{k} \propto R_{ad}^{-2}$. Therefore the mean speed $\bar{v}\propto R_{ad}^{-1}$, as shown in Figure~\ref{Speed}b.

\section{Triplets in the simulation}

In Eq.~(3) we defined the approximate potential energy of two midges in order to define a pair in the simulation as a bound pair when
\be
\mid E_{k}/U_{\mbox{\scriptsize pair}} \mid <1.
\ee
We used this criterion in order to identify pairs in the simulation. In a similar way we can look for triplets where we have three midges in mutual interactions, but there is one additional subtlety. Suppose we have three midges $a$, $b$ and $c$. Then we could have a situation where $a$ is in a pair with $b$, $b$ is in a pair with $c$, but $a$ is not in a pair with $c$. We keep this case as a triplet in our analysis.

\begin{figure}[htb]
          \includegraphics[width=0.6\linewidth]{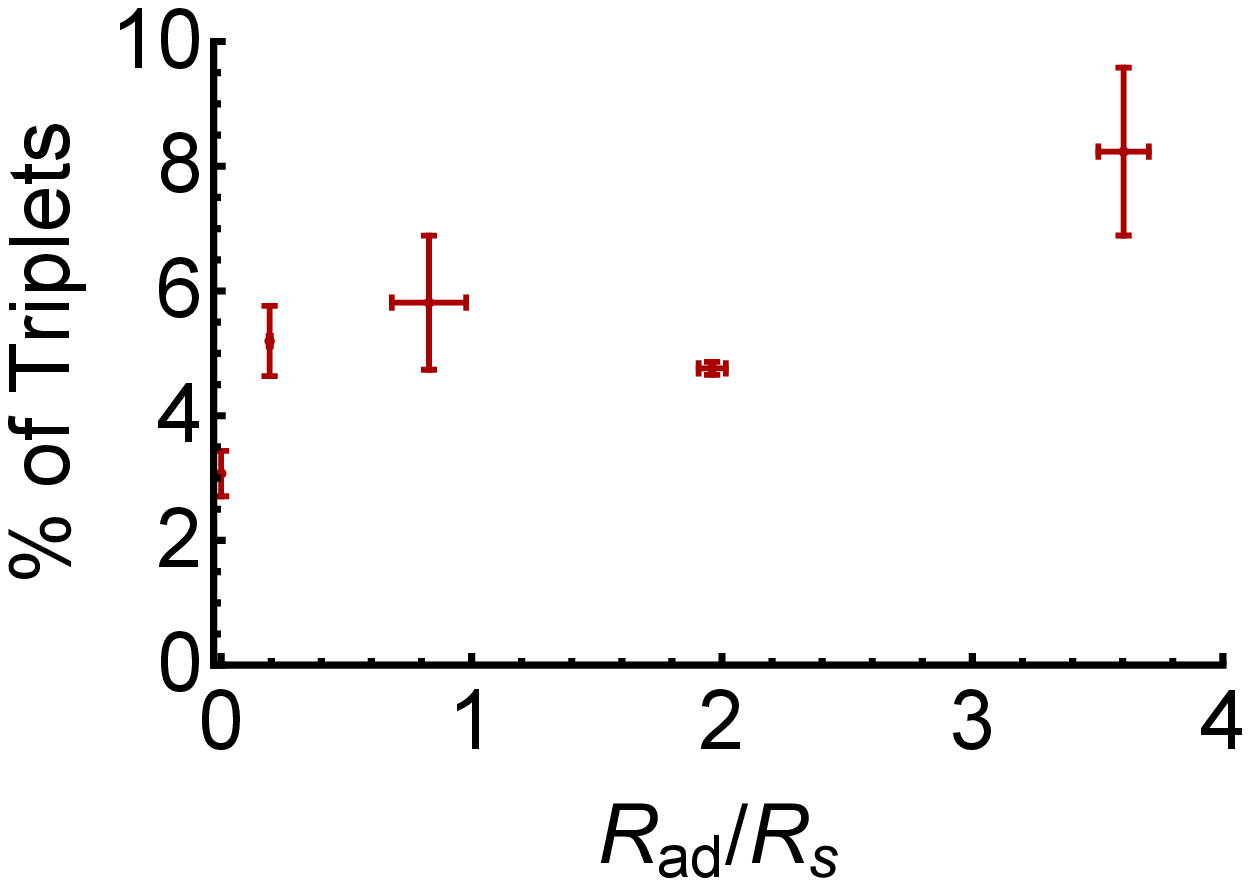}
\vspace{1.5cm}
\caption{The percentage of triplets in the simulation as a function of $R_{ad}/R_{s}$. The number of midges is $N=30$.
\label{triplets}}
 \includegraphics[width=0.6\linewidth]{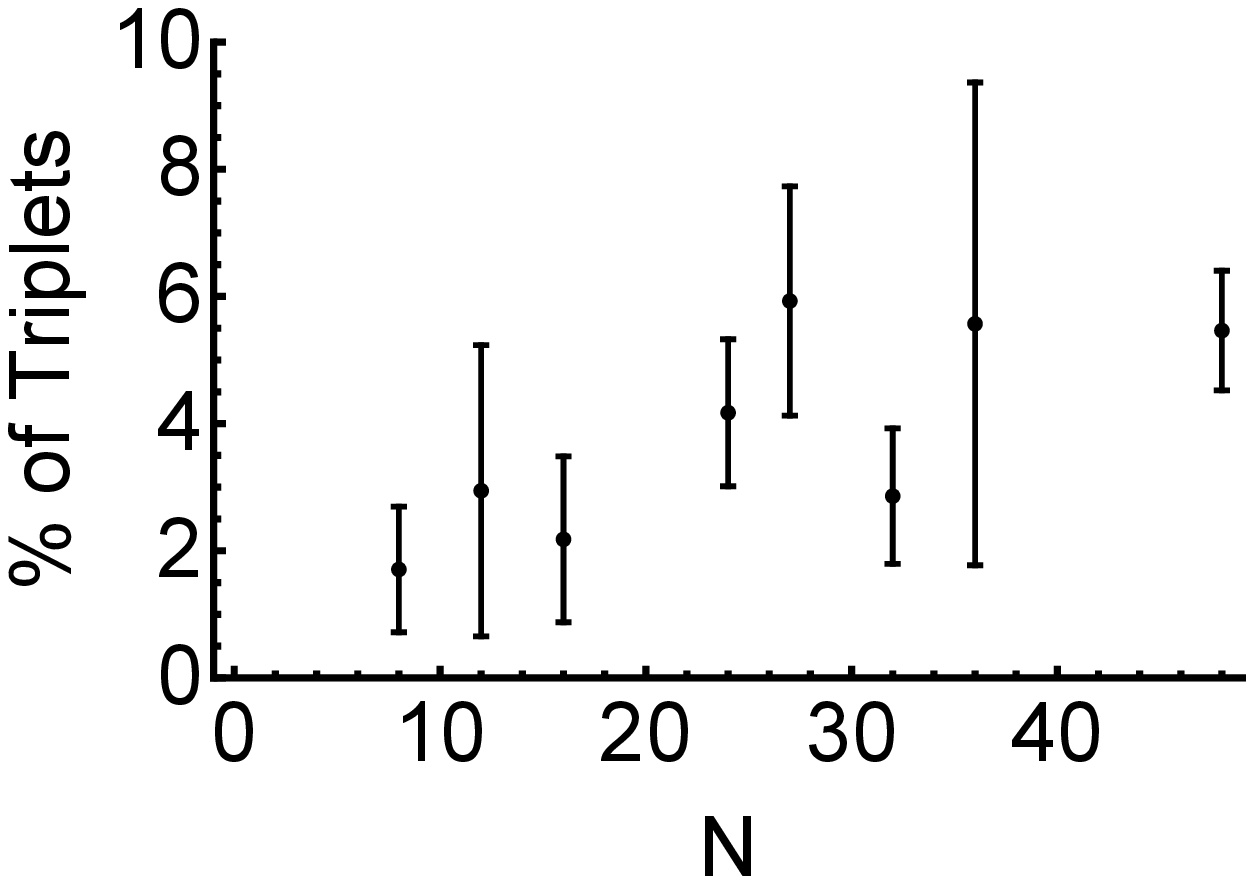}
\caption{  The percentage of pairs as a function of the number of members of the swarm $N$. Here $R_{ad}=10$, and we used various values of $R_s$.
\label{tripletsB}}

\end{figure}

In Fig.~\ref{triplets} we plot the fraction of time a particle spends in a triplet as function of the length scale of adaptivity $R_{\mbox{\scriptsize ad}}$ (for a swarm of $N=30$). It looks that the fraction reaches a plateau around $4-5\%$ which is similar to the plateau of the pairs in Figure 2a, though the result for triplets is more noisy since it is a smaller quantity. The results for triplets from the laboratory data are still too noisy for a reliable comparison with the theoretical ones. In Fig.~\ref{tripletsB} we plot the fraction of triplets as function of the swarm size (for fixed $R_{\mbox{\scriptsize ad}}=10$).

\section{Creation of bound pairs: Analytical approach}
  The effective force felt by one midge due to a second was written in the following way in the Supplementary Information (SI) of~\cite{gorbonos2016long} (as it simply follows Eq.~(1) when we take $\epsilon=0$)):
\begin{equation}
\vec{F}_{\mbox{\scriptsize eff}}=\hat{r}_{12}\frac{C}{|\vec{r}_1-\vec{r}_2|^2}\mathcal{A}(|\vec{r}_1-\vec{r}_2|) \label{feff}
\end{equation}
where $\hat{r}_{12}=(\vec{r}_1-\vec{r}_2)/|\vec{r}_1-\vec{r}_2|$ is the unit vector connecting the two midges and the adaptivity function is given by
\begin{equation}
\mathcal{A}(|\vec{r}_1-\vec{r}_2|)=\frac{R_{\mbox{\scriptsize ad}}^{-2}}{R_{\mbox{\scriptsize ad}}^{-2}+|\vec{r}_1-\vec{r}_2|^{-2}}. \label{adapt}
\end{equation}
$R_{\mbox{\scriptsize ad}}$ is a measure of the maximal distance between midges over which the adaptivity of the midge acoustic sensing can function.
This force can be integrated to give the effective midge-midge potential
\begin{equation}
U_{\mbox{\scriptsize pair}}(|\vec{r}_1-\vec{r}_2|)=\frac{C}{R_{\mbox{\scriptsize ad}}}\left(\arctan{[|\vec{r}_1-\vec{r}_2|/R_{\mbox{\scriptsize ad}}]}-\pi/2\right).
\label{ueff}
\end{equation}


Due to the adaptive interactions (Eq.~(\ref{feff})), when two midges happen by chance to come very close to each other, they
may form a bound pair that is effectively screened from the rest of the swarm. Let us consider the interaction of such a pair in the background provided by the rest. We assume that the separation between the two midges is small compared to their distance to the rest of the swarm, and therefore the interactions with the rest of the swarm will be negligible except for a contribution to the adaptivity factor in Eq.~(\ref{feff}), so that the effective force felt by one member of the pair is
\begin{equation}
\vec{F}_{\mbox{\scriptsize tot},1}\simeq C\hat{r}_{12}\frac{1}{|\vec{r}_1-\vec{r}_2|^2}\left(\frac{R_{\mbox{\scriptsize ad}}^{-2}}{R_{\mbox{\scriptsize ad}}^{-2}+I_{\mbox{\scriptsize background}}+|\vec{r}_1-\vec{r}_{2}|^{-2}}\right), \label{feffmanynn}
\end{equation}
where
\be \label{bgd}
I_{\mbox{\scriptsize background}}=\sum_{i=3}^{n}\frac{1}{|\vec{r}_1-\vec{r}_{i}|^{2}}\sim I_{bgd}+\mathcal{O}(|\vec{r}_1-\vec{r}_{i}|).
\ee
The leading order in the expansion at infinity gives us a constant background sound.

Integrating this force we can calculate the effective two-body potential to be (see Eq.~(\ref{ueff}))
\be
U_{\mbox{\scriptsize pair}}(r)=\frac{C}{\gamma\,R_{\mbox{\scriptsize ad}}}\(\arctan\(\frac{\gamma\,r}{R_{\mbox{\scriptsize ad}}}\)-\frac{\pi}{2}\),
\ee
where $r\equiv|\vec{r}_1-\vec{r}_2|$ and
\be
\gamma\equiv\sqrt{1+R_{ad}^2\,I_{bgd}}
\ee
is the background sound parameter.
We thus effectively have two-body motion under the influence of a mutual central force. Note that in the case of two bodies, additivity of the effective force is valid and as a result we can use all the conservation laws of a central force (energy and angular momentum conservation).
 Therefore this two-body system can be reduced to an equivalent one-dimensional motion in the effective potential
\be \label{gen_potential}
U_{\mbox{eff},12}(r)=\frac{\tilde{l}^2}{2\,r^2}+\frac{C}{\gamma\,R_{\mbox{\scriptsize ad}}}\(\arctan\(\frac{\gamma\,r}{R_{\mbox{\scriptsize ad}}}\)-\frac{\pi}{2}\),
\ee
where $\tilde{l}$ is the angular momentum per unit mass (where we take the reduced mass of the pair).
Let us consider a simple limit that will make the calculations more transparent and later we will show that in the general case the difference is only quantitative. In this limit we take
$R_{ad} \rightarrow 0$ and $I_{bgd} \rightarrow \infty$, so that $R_{ad}^2\,I_{bgd}$ is finite. The potential energy in this limit is:
\be
U_{\mbox{eff},12}(r)=\frac{\tilde{l}^2}{2\,r^2}-\frac{C}{\gamma^2\,r}+\mathcal{O}(r^{-3})
\ee
\begin{figure}[htb]
     \includegraphics[width=0.3\linewidth]{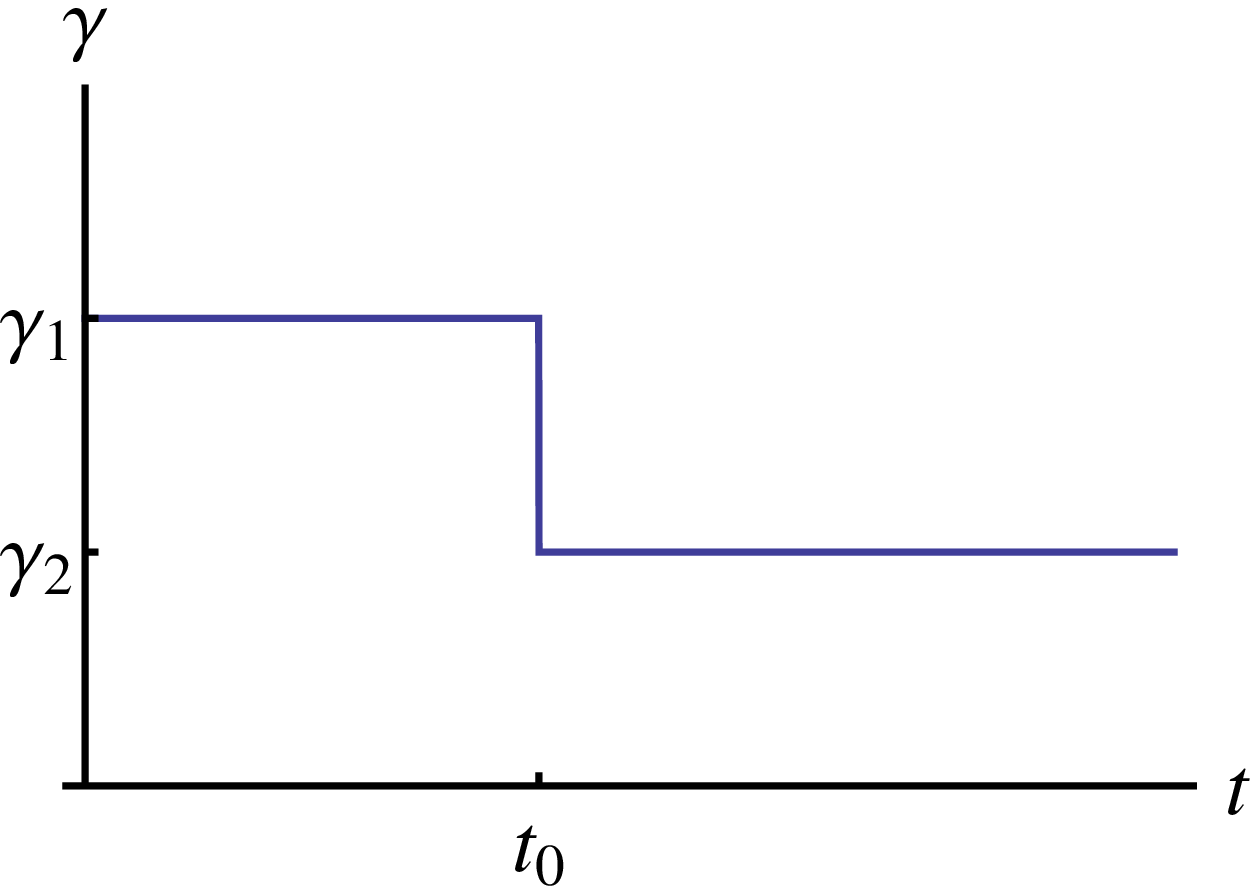}
     \vspace{1.5cm}
\caption{The background sound parameter $\gamma$ is taken to be a step function. For $t\leq t_0$ the background sound level is higher than the background sound for $t> t_0$.
}
\label{NoisePlot}
\end{figure}
When the background sound is reduced, the sensitivity of each member of a pair to its companion becomes higher. As a result their effective mutual attractive force becomes stronger, and this way a pair could form.
Let us analyze it for simplicity when the background sound parameter is a step function (Fig.~\ref{NoisePlot}):
\be
\gamma=\begin{cases}
      \gamma_1 & t  \leq t_0 \\
      \gamma_2 & t > t_0,
   \end{cases}
   \label{step}
\ee
where $\gamma_1>\gamma_2$.

When the mutual attractive force is stronger, even when we start from an elliptical bound orbit, it becomes tighter and therefore the particles approach closer to each other. The total energy, which is conserved, is given by the following expression:
\be \label{energy}
E_{i}=E_{k}(r)+\frac{\tilde{l}^2}{2\,r^2}-\frac{C}{\gamma_{i}^2\,r}, \quad i=1,2
\ee
where $E_{k}(r)$ is the kinetic energy of the radial movement.
At $t=t_0$ let us denote the radial distance of the reduced mass by $r_0$ and the energy of the pair is changing due to the change in the background sound.
The energy difference is given by:
\be \label{DifferenceEnergy}
\Delta E = \Delta U_{\mbox{eff},12}=\(\frac{1}{\gamma_1^2}-\frac{1}{\gamma_2^2}\)\frac{C}{r_0}<0.
\ee
The ``tightness'' of an orbit can be regarded as the ratio of the kinetic to potential energy $\left|E_k/ U_{\mbox{eff},12}\right|$. When it smaller than one $\left|E_k/ U_{\mbox{eff},12}\right|<1$, the orbit is bound. Since the effective potential energy is lowered as a result of the reduction of the background sound and the kinetic energy remains the same, this process may produce bound orbits out of unbound ones.

As it is known for Keplerian orbits, when the energy or angular momentum become too high, the orbits are unbound. In order to characterize the range of parameters for which the orbits are bound (ellipses for Keplerian orbits) it is convenient to introduce two dimensionless quantities that correspond to the energy and angular momentum squared:
\bea \label{definitions}
\epsilon\equiv\frac{R_{ad}\,|E|}{C} \nonumber\\
j^2\equiv\frac{\tilde{l}^2}{2\,R_{ad}\,C}
\eea
Then the orbits are bound if there are solutions to the following quadratic equation:
\be \label{quadratic}
\gamma^2\,\epsilon\,\overline{r}^2-\overline{r}+\gamma^2\,j^2=0,
\ee
where $\overline{r}\equiv r/R_{ad}$ and whose solutions give the periapsis and apoapsis of the elliptical orbit.
Such solutions exist if and only if
\be
\epsilon\,j^2<\frac{1}{4\,\gamma^4}.
\ee
When $\gamma$ is lower (the background sound is reduced), the bound orbits exist for more possible values of $\epsilon$ and $j^2$. In particular, values that did not correspond to a bound orbit now produce such an orbit.

In addition, when $\gamma$ is lower, the apoapsis of the bound (elliptical) orbit is reduced and the two particles move closer to each other during their orbit (one around the other). The maximal distance between them is given by
\be \label{rmax}
\overline{r}_{max}(\gamma)=\frac{2\,\gamma^2\,j^2}{1-\sqrt{1-4\gamma^4\,\epsilon(\gamma)\,j^2}}.
\ee
In order to show the reduction of the maximal distance explicitly let us look at a small change in the background sound:
\be
\gamma=\gamma_1+\delta.
\ee
at $t=t_0$ when the particle is at $\overline{r}=\overline{r}_0$. Then the (dimensionless) energy changes according to:
\be
\epsilon(\gamma)=\epsilon(\gamma_1)+\left.\frac{d\epsilon}{d\delta}\right|_{\delta=0}\!\!\!\!\delta=\epsilon(\gamma_1)-\frac{2}{\gamma_1^3\overline{r}_0}\delta+\mathcal{O}(\delta^2),
\ee
and the maximal distance is then
\be
\overline{r}_{max}(\gamma)=\overline{r}_{max}(\gamma_1)+\left.\frac{d\overline{r}_{max}}{d\delta}\right|_{\delta=0}\!\!\!\!\delta+\mathcal{O}(\delta^2),
\ee
where
\be
\left.\frac{d\overline{r}_{max}}{d\delta}\right|_{\delta=0}=\frac{4\,\gamma_1\,j^2}{\(1-\sqrt{1-4\gamma^4\,\epsilon\,j^2}\)^2}\(1-\frac{1}{\sqrt{1-4\gamma^4\,\epsilon\,j^2}}+\frac{2\,\gamma_1^2\,j^2}{\overline{r}_0\,\sqrt{1-4\gamma^4\,\epsilon\,j^2}}\).
\ee
Since $\overline{r}_0<\overline{r}_{max}$, which is given in Eq.~(\ref{rmax}), the expression in the parentheses is always positive and thus
\be
\left.\frac{d\overline{r}_{max}}{d\delta}\right|_{\delta=0}>0. \label{rmaxder}
\ee
To conclude, from the inequality~(\ref{rmaxder}) we see that when the background sound is higher, the maximal distance is higher and when the background sound is lower, the orbit is tighter and the maximal distance is reduced.

\begin{figure}[htb]
     \includegraphics[width=0.6\linewidth]{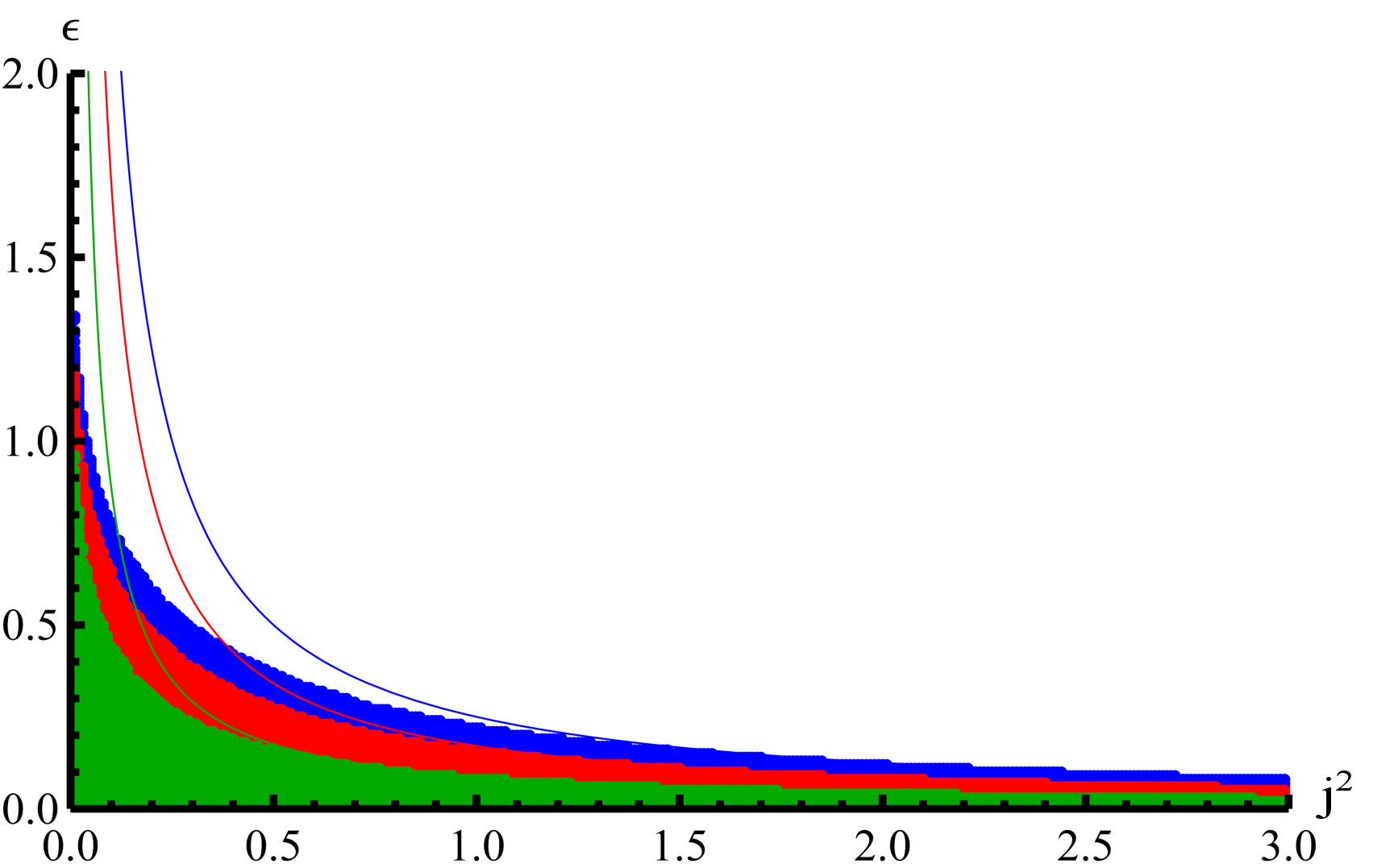}
     \vspace{1.5cm}
\caption{Points in the $j^2-\epsilon$ parameter space (dimensionless angular momentum squared and dimensionless energy) for which there are bound orbits in the effective potential. The blue, red and green regions correspond to $\gamma=1,1.1,1.3$ respectively. The lines correspond to $1/(4\,\gamma^4)$ curves.}
\label{gammaValues}
\end{figure}

Let us now consider a more general case where $R_{ad}$ has an arbitrary value and use the same step function for the background sound~(\ref{step}). In this case the total energy is given by the following expression:
\be
\label{energy_gen}
E_{i}=E_{k}(r)+\frac{\tilde{l}^2}{2\,r^2}+\frac{C}{\gamma_{i}\,R_{\mbox{\scriptsize ad}}}\(\arctan\(\frac{\gamma_{i}\,r}{R_{\mbox{\scriptsize ad}}}\)-\frac{\pi}{2}\), \quad i=1,2
\ee
 and the energy difference at $t=t_0$ is given by
\be
 \label{DifferenceEnergy_gen}
\Delta E = \Delta U_{\mbox{eff},12}=\frac{C}{\gamma_{2}\,R_{\mbox{\scriptsize ad}}}\(\arctan\(\frac{\gamma_{2}\,r_0}{R_{\mbox{\scriptsize ad}}}\)-\frac{\pi}{2}\)-\frac{C}{\gamma_{1}\,R_{\mbox{\scriptsize ad}}}\(\arctan\(\frac{\gamma_{1}\,r_0}{R_{\mbox{\scriptsize ad}}}\)-\frac{\pi}{2}\).
\ee
When $\gamma_2<\gamma_1$ in this case $\Delta E <0$ since $U_{\mbox{eff},12}(r)$ which is given in Eq.~(\ref{gen_potential}) is monotonically increasing as a function of $\gamma$ for any positive value of $R_{ad}$:
\be
\partial_{\gamma}U_{\mbox{eff},12}=C\,\frac{2\,\gamma\,R_{ad}\,r+(R_{ad}^2+\gamma^2\,r^2)(\pi-2\,\arctan\(\frac{\gamma\,r}{R_{ad}}\))}{2\,\gamma^2\,R_{ad}\(R_{ad}^2+\gamma^2\,r^2\)}>0.
\ee
Let us consider bound orbits in the Hamiltonian system that is given by Eq.~(\ref{energy_gen}). Again, as previously discussed, we want to find the values of $\epsilon$ and $j^2$ that correspond to bound orbits. Taking $E_{k}(r)=0$ at the extremal points of the orbit we get the following equation for $\overline{r}$:
\be
\label{rbar}
{\overline{r}}^2\,\epsilon+j^2+\frac{\overline{r}^2}{\gamma}\(\arctan\(\overline{r}\gamma\)-\frac{\pi}{2}\)=0.
\ee
In the limit $R_{ad}\rightarrow0$ we recover Eq.~(\ref{quadratic}) as expected.
This equation can be solved numerically for different values of $\epsilon$ and $j^2$.
In Fig.~(\ref{gammaValues}) we give three examples for different values of $\gamma$ ($\gamma=1,1.1,1.3$). The colored region correspond to values in the $j^2-\epsilon$ plane, to which there are solutions of Eq.~(\ref{rbar}). Each region consists of some of the colored patches, so that the largest region is the one that corresponds to $\gamma=1$ which consists of the blue, red and green patches. $\gamma=1.1$ corresponds to the red and the green patches and $\gamma=1.3$ corresponds only to the green patch. Therefore in general when the background sound is reduced, we have a larger region in the parameter space for bound orbits. The colored curves that appear are the corresponding $1/(4\,\gamma^4)$ graphs in the $R_{ad}\rightarrow 0$ approximation. In this approximation $\epsilon\rightarrow 0$ and $j^2 \rightarrow \infty$ while $\epsilon j^2$ is finite. That is why in the limit of $\epsilon\rightarrow 0$ and $j^2 \rightarrow \infty$ the curves $1/(4\,\gamma^4)$ are an asymptote to the region of bound orbits.
We cannot write an explicit expression for $\overline{r}_{max}$ in the general case but the case when $j=0$ is easily solvable. The solution to Eq.~(\ref{rbar}) is
\be \label{rmaxj0}
\overline{r}_{max}=\frac{1}{\gamma}\tan(\frac{\pi}{2}-\gamma\epsilon(\gamma)),
\ee
Eq.~(\ref{energy_gen}) can be rewritten at the point $\overline{r}=\overline{r}_0$ (where we change the background sound $\gamma$):
\be \label{epsbar}
\epsilon=-\epsilon_{k}-\frac{j^2}{\overline{r}_0^2}+\frac{1}{\gamma}\(\frac{\pi}{2}-\arctan(\gamma\,\overline{r}_0)\),
\ee
where
\be
\epsilon_k\equiv\frac{R_{ad}\,E_k}{C}
\ee
and substituting into Eq.~(\ref{rmaxj0}) in the case of $j=0$ gives us
\be
\overline{r}_{max}=\frac{1}{\gamma}\tan(\gamma \,\epsilon_{k}+\arctan(\gamma\,\overline{r}_0)).
\ee
It can be shown explicitly that when $j=0$
\be
\frac{\partial \overline{r}_{max}}{\partial \gamma}>0,\label{rmaxder2}
\ee
since we can write it as a sum of two positive quantities:
\be
\frac{\partial \overline{r}_{max}}{\partial \gamma}=\frac{\overline{r}_0^2\,\(2\,\gamma\,\epsilon_{k}+\sin(2\,\gamma\,\epsilon_k)\)}{2\left[\cos(\gamma\,\epsilon_k)-\gamma\,\overline{r}_0\,\sin(\gamma\,\epsilon_k)\right]^2}+\frac{2\,\gamma\left[\epsilon_k+\overline{r}_0-\overline{r}_0\,\cos(2\,\gamma\,\epsilon_k)\right]-\sin(2\,\gamma\,\epsilon_k)}{2\gamma^2\left[\cos(\gamma\,\epsilon_k)-\gamma\,\overline{r}_0\,\sin(\gamma\,\epsilon_k)\right]^2}.
\ee

When $j\neq0$ from Eqs.~(\ref{rbar},\ref{epsbar}) we find
\be \label{CombinedGen}
\frac{j^2}{\overline{r}_{max}^2}-\frac{j^2}{\overline{r}^{2}_{0}}-\epsilon_{k}+\frac{1}{\gamma}\(\arctan(\gamma\,\overline{r}_{max})-\arctan(\gamma\,\overline{r}_0)\)=0
\ee
To look at the maximal distance $\overline{r}$ as a function of the background sound $\gamma$ we can take the derivative of Eq.~(\ref{CombinedGen}) with respect to $\gamma$ and obtain:
\be \label{longFraction}
\frac{\partial \overline{r}_{max}}{\partial \gamma}=\(\arctan(\gamma\,\overline{r}_{max})-\arctan(\gamma\,\overline{r}_0)+\frac{\gamma\,\overline{r}_{0}}{1+\gamma^2\overline{r}_{0}^2}-\frac{\gamma\,\overline{r}_{max}}{1+\gamma^2\overline{r}_{max}^2}\)
\left/\(\frac{\gamma^2}{1+\gamma^2\overline{r}_{max}^2}-\frac{2\,\gamma^2\,j^2}{\overline{r}_{max}^3}\)\right..
\ee
Again we want to check when $\frac{\partial \overline{r}_{max}}{\partial \gamma}>0$, namely the maximal distance is larger when the background sound is higher. The numerator of (\ref{longFraction}) is always positive since $\overline{r}_0<l \overline{r}_{max}$. Therefore the sign of (\ref{longFraction}) is determined by the denominator.
If
\be \label{inequa}
j^2<\frac{1}{2}\,\frac{\overline{r}_{max}^3}{1+\gamma^2\overline{r}_{max}^2}
\ee
the denominator of $(\ref{longFraction})$ is positive, and then
\[\frac{\partial \overline{r}_{max}}{\partial \gamma}>0,\]
as we expect to happen. For $j=0$ this inequality is trivially satisfied and we saw already that the derivative of $\overline{r}_{max}$ with respect to the background sound is positive in Eq.~(\ref{rmaxder2}). In Eq.~(\ref{rmaxder}) we saw that it is true also for the limit when $j\rightarrow\infty$, which corresponds to $R_{ad}\rightarrow 0$ according to the definition of $j^2$ (Eq.~(\ref{definitions})), and it can be confirmed numerically for any positive value of $j^2$ for which a solution to Eq.~(\ref{rbar}) exists.

\section{The three body problem}

In order to illustrate the crucial role of adaptivity in the creation of a bound pair, it is instructive to consider a toy model of three interacting bodies when one of the bodies is much heavier than the other two and plays the role of the background swarm. The other two bodies play the role of the interacting midges that produce a bound pair. Our goal in this toy model is to look at similar trajectories with and without adaptivity (the $\epsilon$-gravity model) and in this way to observe the influence of the background sound $I_{background}$ (which is produced by the heavy mass in the adaptive case).

\begin{figure}[htb]
     \includegraphics[width=0.6\linewidth]{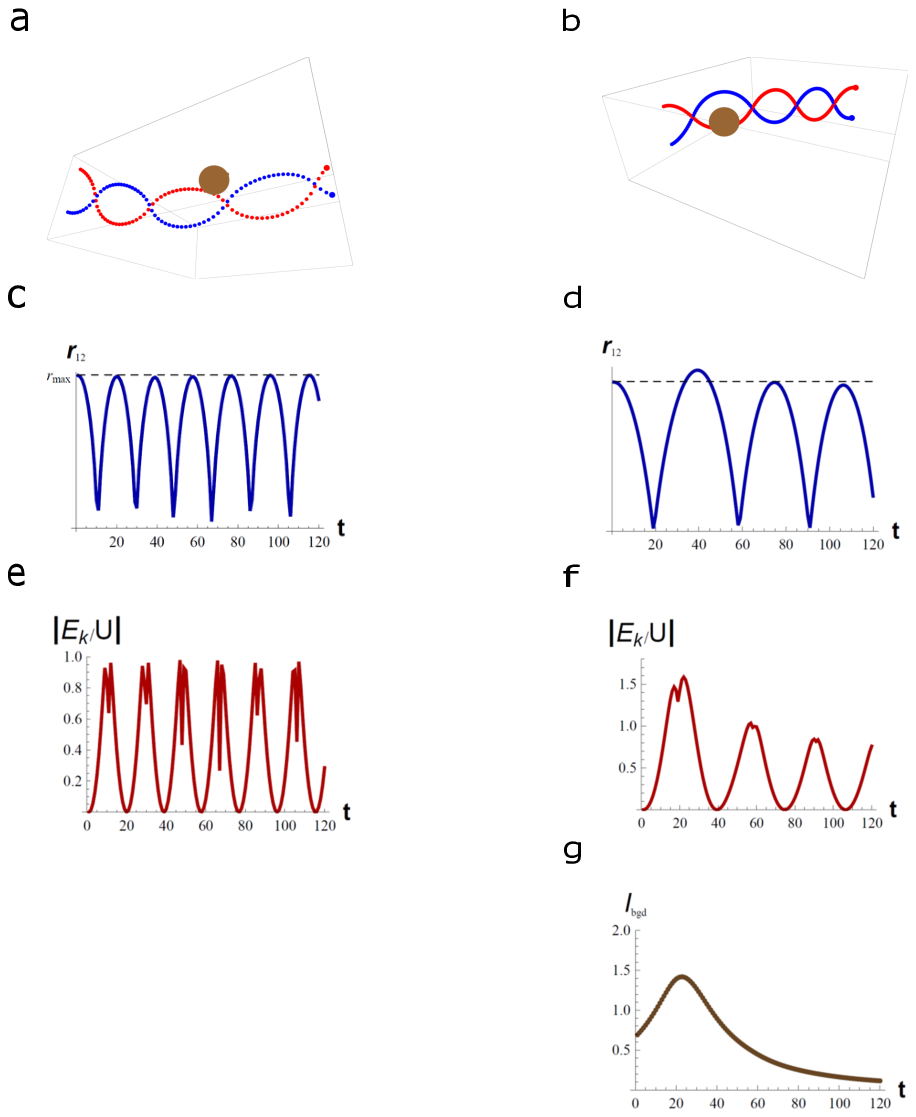}

     \vspace{2cm}
\caption{(a) The trajectories of three bodies without adaptivity (the $\epsilon$-gravity model)- Two identical small bodies pass near a heavy one (b) The same as (a) but with adaptivity ($R_{ad}=5$) (c) The relative distance between the pair of the small masses in (a) as a function of simulation time.  (d) The same for the pair in (b). (e) The ratio of the kinetic to potential energy as a function of simulation time for one of the small masses in (a). (f) The same ratio but for the trajectory in (b). (g) The ``background sound'' as a function of time which was defined in Eq.~(\ref{bgd}) $I_{background}$.
}
\label{3bodies}
\end{figure}

We start in a configuration without relative velocity in the pair but where both have a velocity relative to the heavy body (the trajectory is shown in Fig~\ref{3bodies}a,b). We take in this example $C=1$, $\epsilon^2=1.5$ and $R_{ad}=5$ in Eq.~(1) for the case of three bodies and obtain the trajectory in Fig~\ref{3bodies}b with the initial conditions:
\bea
\vec{r}_{1}&=&(20,5,-2.5) \quad \vec{v}_1=(-1,0,0) \nonumber\\
\vec{r}_{2}&=&(20,5,2.5) \quad \vec{v}_2=(-1,0,0) \nonumber\\
\vec{r}_{3}&=&(0,25,0) \quad \vec{v}_3=(0,0,0)
\eea
where $\vec{r}_{1}$, $\vec{r}_{2}$, $\vec{v}_{1}$, and $\vec{v}_{2}$ are the initial positions and velocities of the unit mass bodies and the third body (whose initial position and velocity are $r_{3}$ and $v_{3}$) has a mass that is 30 times higher. In order to create similar trajectories without adaptivity ($\epsilon$-gravity), we have to take smaller masses and larger distances since the adaptive force is weaker by a factor of $R_{ad}^2$. The trajectories in Fig.~\ref{3bodies}a are produced by taking
\be
\vec{r}_{3}=(0,40,0) \quad \vec{v}_3=(0,0,0)
\ee
and a mass that is 25 times higher than the small unit masses.

When we compare the relative distance between the pair of the small masses as a function of simulation time (Fig.~\ref{3bodies}c,d), we see that without adaptivity the heavy mass has no influence on the relative distance between the pair and their orbits, one with respect to the other, are unchanged. Throughout the trajectory the bound pair remains bound, as seen in their ratio of relative kinetic and potential energies (Fig.~\ref{3bodies}e). In the case with adaptivity, we see that when the pair is closest to the heavy mass, the pair tend to separate - their relative distance is maximal. This is the result of the strong ``background sound'' that lowers their attraction to each other. At this point they are not a bound pair anymore, as seen in the ratio of their relative kinetic and potential energies, which is larger than 1 (Fig.~\ref{3bodies}f). When the pair leaves the influence of the heavy mass, their mutual attractive force becomes stronger again and a bound pair is formed, as we see from the ratio of kinetic to potential energies (Fig.~\ref{3bodies}f). The increase and decrease in the ``background sound'' along the trajectory is given in Fig.~\ref{3bodies}g. As described in this article, this mechanism is the dominant producer of bound pairs in the simulation.

\section{Spatial Distribution of Pairs (simulation)}

\begin{figure}[htb]
     \includegraphics[width=0.6\linewidth]{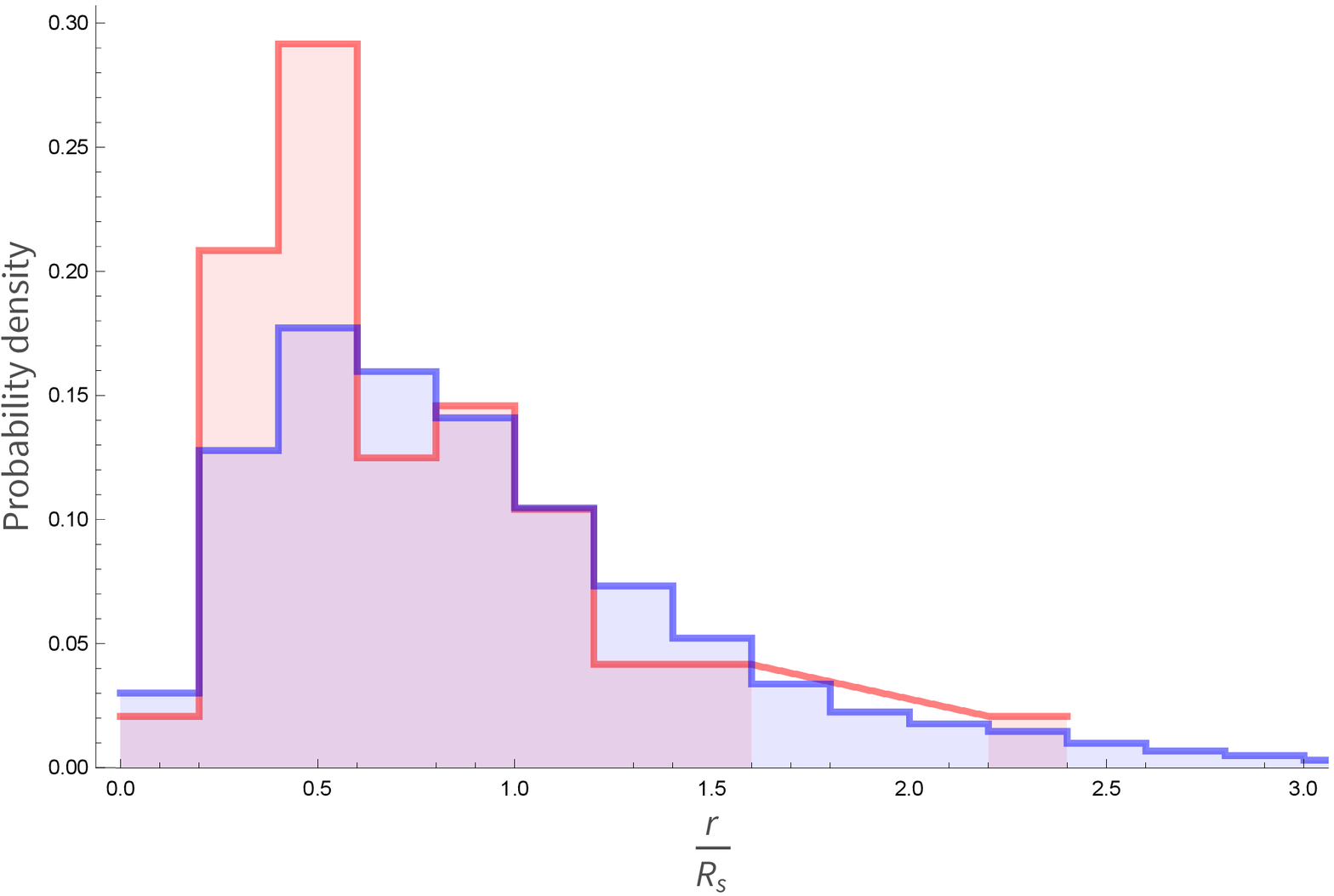}

     \vspace{2cm}
\caption{The spatial distribution of pairs at the moment of pair formation (simulation for $N=36$, $R_{ad}=50$). In red the PDF of the distance of the pairs from the center of mass. In blue, for comparison, the PDF of the distance of all the midges in the swarm. Both are normalized with respect to the mean radius of the swarm $R_{s}$.
}
\label{pairsSpatial}
\end{figure}

In Fig.~\ref{pairsSpatial} we present the radial distribution of pair-formation events, in comparison with the radial Gaussian distribution of particles in the same simulated swarm. We can see that the pairs are usually formed in an intermediate density region ($\sim 0.5 R_{s}$).

We can estimate the spatial distribution of the pair formation rate using the following relation:
\be
{\dot\rho}_{pair}(r)\sim k(r) \rho(r)^2,
\label{kpair}
\ee
where ${\dot\rho}_{pair}$ is the local rate (probability per unit time) to form a new pair at $r$, $k(r)$ is the trapping rate per midge and $\rho(r)$ is the density of midges.

\begin{figure}[htb]
\includegraphics[width=0.6\linewidth]{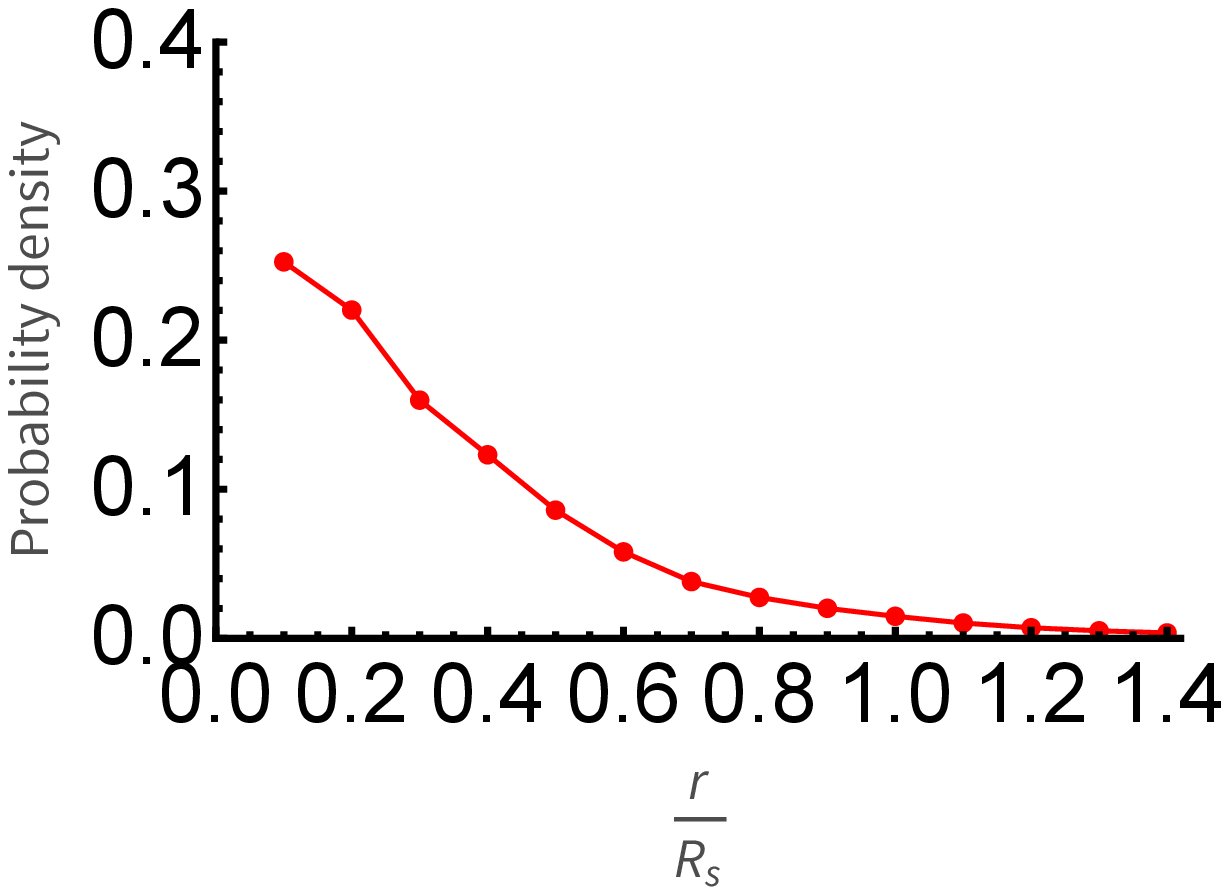}

     \vspace{2cm}
\caption{The general density of midges $\rho(r)$ in the central region ($r<R_{s}$). (Normalized as PDF for $r<R_{s}$.)}
\label{densityhist}
\end{figure}

\begin{figure}[htb]
\includegraphics[width=0.6\linewidth]{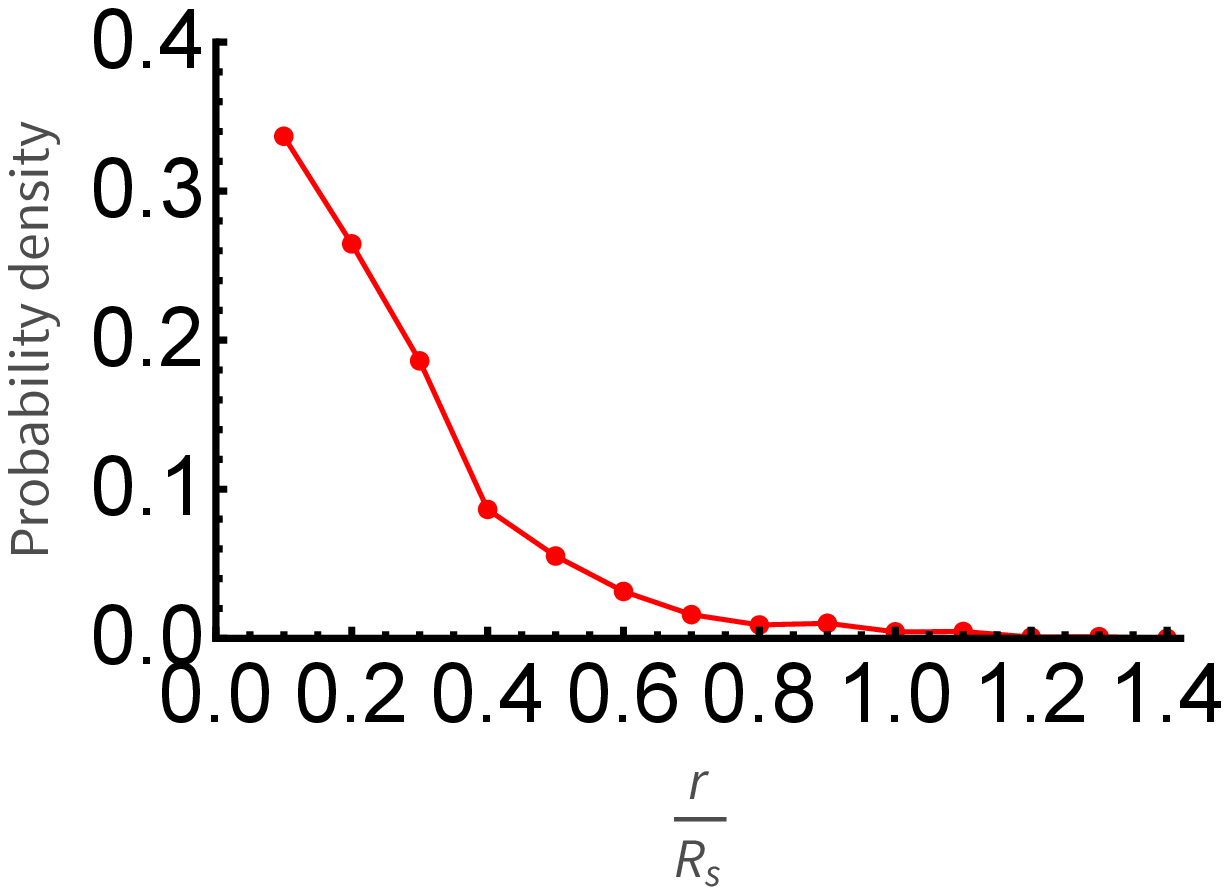}

     \vspace{2cm}
\caption{The density of pair formation events ${\dot\rho}_{pair}$ in the central region ($r<R_{s}$). (Normalized as PDF for $r<R_{s}$.)}
\label{pairhist}
\end{figure}

\begin{figure}[htb]
\includegraphics[width=0.6\linewidth]{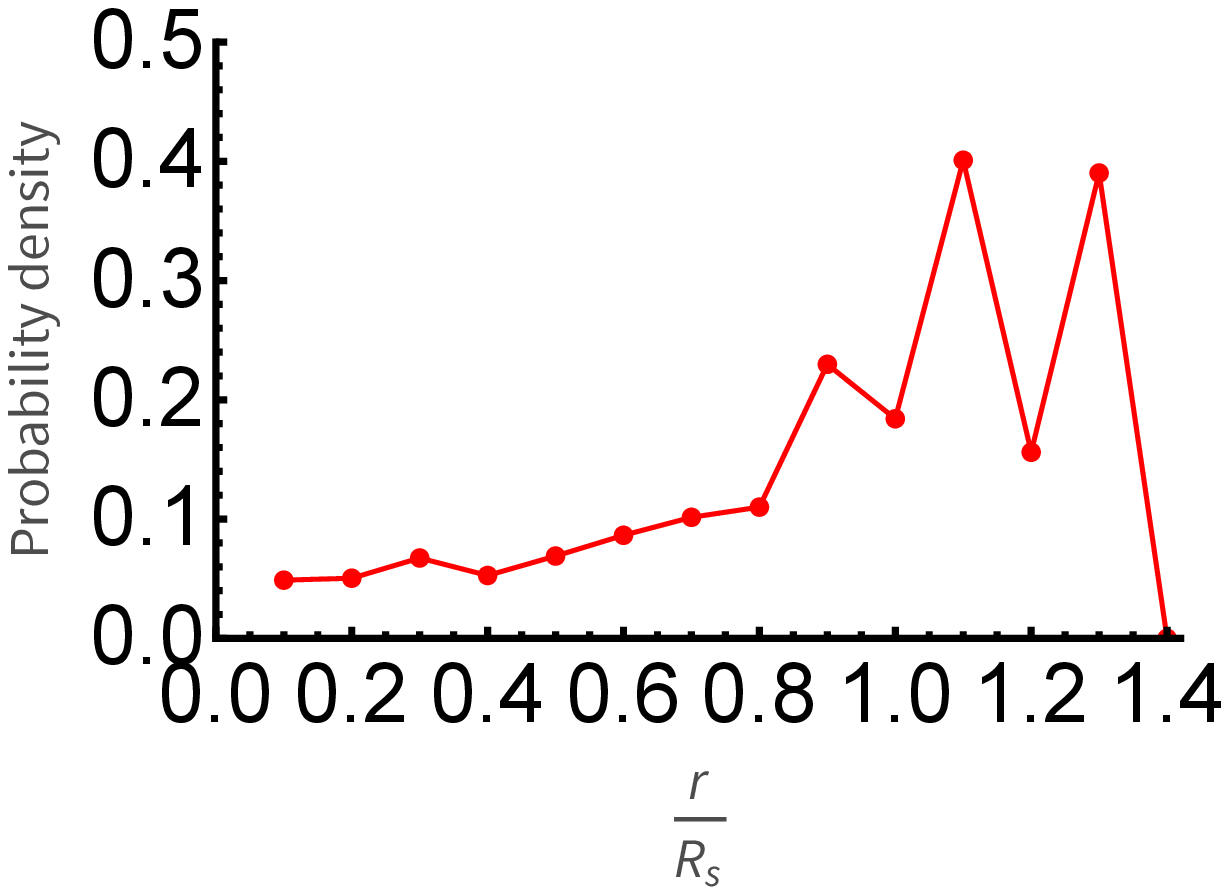}
     \vspace{2cm}
\caption{The rate of pair formation per midge, $k(r)$, in the central region ($r<R_{s}$), from Eq.\ref{kpair}. (Normalized as PDF for $r<R_{s}$.)}
\label{kvalhist}
\end{figure}

Let us estimate the pair formation rate $k(r)$ for the swarm that was considered in Fig.~\ref{pairsSpatial}. The radial density $\rho(r)$ is given in Fig.~\ref{densityhist}, the radial density of pair-formation events (over the course of a given simulation time, so proportional to ${\dot\rho}_{pair}(r)$) is in Fig.~\ref{pairhist}, and from them we get the rate of pair formation per midge $k(r)$ in Fig.~\ref{kvalhist}. There is a peak in the pair formation rate per midge at $r \sim 0.9 R_{s}$.

\begin{figure}[htb]
\includegraphics[width=0.6\linewidth]{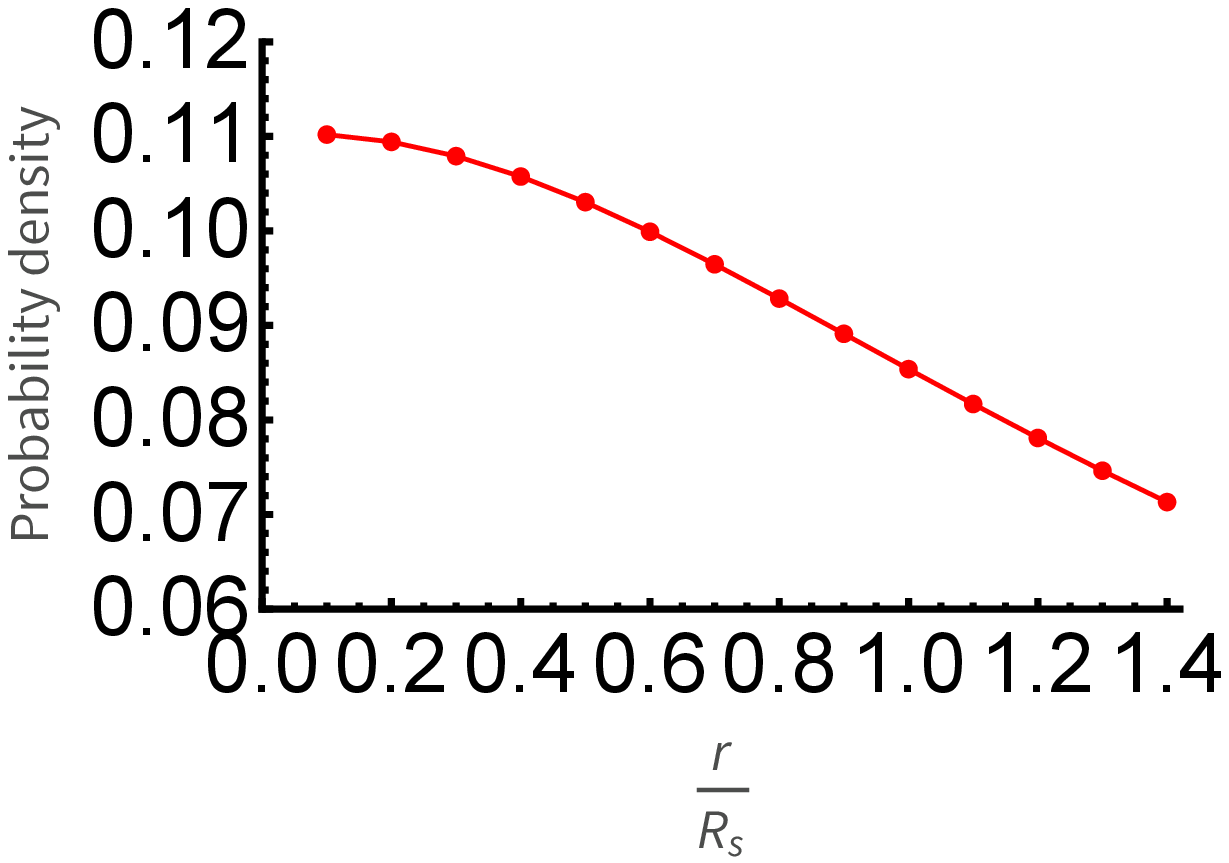}
     \vspace{2cm}
\caption{The total background sound in the central region ($r<R_{s}$). (Normalized as PDF for $r<R_{s}$.)}
\label{BgdSoundhist}
\end{figure}

\begin{figure}[htb]
\includegraphics[width=0.6\linewidth]{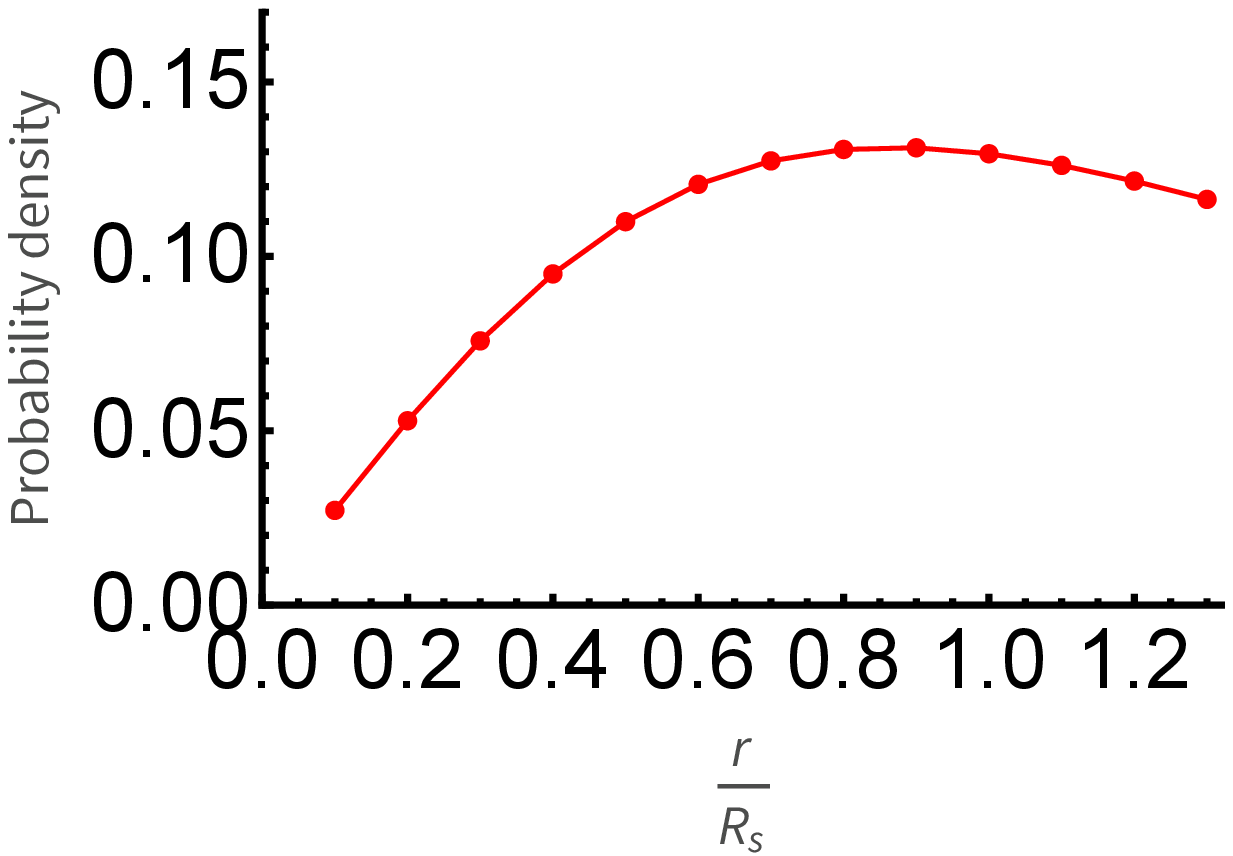}
     \vspace{2cm}
\caption{Estimation of the radial gradient in the background sound, in the central region ($r<R_{s}$). (Normalized as PDF for $r<R_{s}$.)}
\label{derGammahist}
\end{figure}

According to the adaptive-gravity model that we described, there should be a correlation between the pair formation rate and the gradient of the total background sound (see Eqs.~(3-5)). The mean total background sound was computed from the simulation and is given in Fig.~\ref{BgdSoundhist}. It is very close to the distribution of background sound for a Gaussian distribution of particles with the same $R_{s}$. The estimation of the radial derivative of the total sound is given in Fig.~\ref{derGammahist}. Indeed, we find that the peak of the background sound gradient is at $r\sim 0.9 R_s$, which is close to the peak location in $k(r)$.

All the results of this section are based on simulation of 36 particles for $\epsilon^2=15$ and $R_{ad}=50$.

\bibliography{acoustics_1}